\documentclass[12]{article}
\usepackage{graphics}
\newcommand{\beq}{\begin{equation}}
\newcommand{\eeq}{\end{equation}}

\newcommand{\beqs}{\begin{eqnarray}}
\newcommand{\eeqs}{\end{eqnarray}}

\begin{document}
\bibliographystyle{h-physrev}
\begin{titlepage}
\vskip 2.5cm
\begin{center}
{\LARGE Mass-Gaps and Spin Chains for (Super) Membranes}\\
\vskip 1.2cm
{\Large Abhishek Agarwal} \
\vskip 1.3cm {\Large Physics Department}\\
{\large City College of the CUNY}\\
{\large New York, NY 10031}\\
abhishek@sci.ccny.cuny.edu \vskip 0.7cm
\end{center}
\vspace{3.14cm}
\begin{abstract}
We present a  method for computing the non-perturbative mass-gap in
the theory of Bosonic membranes in flat background spacetimes. The
analysis is extended to the study of membranes coupled to background
fluxes as well. The computation of mass-gaps is carried out using a
matrix regularization of the membrane Hamiltonians. The mass gap is
shown to be naturally organized as an expansion in a 'hidden'
parameter, which turns out to be $\frac{1}{d}$: d being the related
to the dimensionality of the background space. We then proceed to
develop a large $N$ perturbation theory for the
membrane/matrix-model Hamiltonians around the quantum/mass corrected
effective potential. The same parameter that controls the
perturbation theory for the mass gap is also shown to control the
Hamiltonian perturbation theory around the effective potential. The
large $N$ perturbation theory is then translated into the language
of quantum spin chains and the one loop spectra of various Bosonic
matrix models are computed by applying the Bethe ansatz to the
one-loop effective Hamiltonians for membranes in flat space times.
The spin chains corresponding to the large $N$ effective
Hamiltonians for the relevant matrix models are generically not
integrable. However, we are able to find large integrable
sub-sectors for all the spin chains  of interest. Moreover, the
continuum limits of the spin chains are mapped to integrable
Landau-Lifschitz models even if the underlying spin chains are not
integrable.  Apart from membranes in flat spacetimes, the recently
proposed matrix models (hep-th/0607005) for non-critical membranes
in plane wave type spacetimes are also analyzed within the paradigm
of quantum spin chains. The Bosonic sectors of all the models
proposed in (hep-th/0607005) are diagonalized at the one-loop level
and an intriguing connection between the existence of supersymmetric
vacua and one-loop integrability is also presented.
\end{abstract}

\end{titlepage}
\section{Introduction and Summary}
In the present paper we compute the one-loop large $N$ spectrum of
various models of matrix quantum mechanics describing the motion of
membranes with and without supersymmetry. The purely Bosonic cases
discussed in the paper correspond to spherical membranes in flat
backgrounds which we also generalize to include background fluxes.
We also study the one loop spectrum of the Bosonic sectors of
various supersymmetric matrix models that were recently proposed by
Kim and Park\cite{kp} as the regularized descriptions of
supermembranes in non-critical super gravity theories in the
background of plane wave like curved spacetimes. The matrix models
corresponding to the supersymmetric membranes have explicit
quadratic 'mass' terms in their Hamiltonians and their spectra are
clearly discrete. This allows us to study their large $N$ spectrum
by  Bethe ansatz methods which have been very successful in the
study of the the BMN matrix model. However, this is not the
situation for the non-supersymmetric cases that we study. Since the
background spacetime for membrane motion is chosen to be flat, the
membrane Hamiltonians do not have mass terms and are ostensibly
plagued with the problem of the existence of classical flat
directions. Thus a straightforward application of the spin chain
techniques to these models of membrane dynamics is not possible.
However, the problem of existence of flat directions for
non-supersymmetric membranes is entirely an artifact of choosing a
particularly difficult classical theory as a starting point for
quantization. The flat directions are indeed lifted upon
quantization\cite{hopth}. To make use of this simplification, we
develop a method for estimating the dynamically generated mass terms
in the quantum theory for purely bosonic membranes. The upshot of
this technique is that one can develop a controlled large $N$
perturbation theory both for estimating the spectral gap of the
theories as well as for quantizing the theories around the 'quantum
corrected' effective potential. This quantization naturally allows
one to apply the techniques of quantum spin chains and the Bethe
ansatz to systematically study the spectrum of non-supersymmetric
membranes.

The close connection between matrix quantum mechanics and theories
of D branes and membranes has been known for quite
sometime\cite{taylor1,taylor2} and one of the principal motivations
leading to the present work is the search for methods that might
help us understand the gauge- theory/gravity correspondence from the
point of view of D branes and other extended objects. A key insight
gained from the recent advances in the AdS/CFT correspondence is
that integrable structures naturally manifest themselves on both the
string and gauge theory sides of the correspondence. As far as the
spectrum of closed string like excitations, both from the gauge
theory as well as the string theoretic point of view, are concerned,
there is by now mounting evidence that they are  described by
integrable systems. On the gauge theory end, one concerns oneself
with the spectrum of anomalous dimensions of single trace local
composite operators. To the extent that it has been possible to
check so far, this spectrum appears to be the spectrum of an
integrable spin chain\cite{mz,bs1}. On the string theory side, the
corresponding integrable system is nothing but the world sheet sigma
model, which too appears to be integrable, at least at the classical
level\cite{bpr,afs,art}. If one focusses on the gauge theory, then
it is not difficult to see that the fundamental system leading to
large $N$ integrability is a quantum mechanical matrix model. This
matrix model is nothing but the dilatation operator of the gauge
theory, which in the large $N$ limit can be interpreted as an
integrable quantum spin chain\cite{ar1,ar2,ber1,bell}. Thus, keeping
in mind that D brane and membrane world volume theories are nothing
but quantum mechanical matrix models, it is perhaps reasonable to
expect that integrable systems such as quantum spin chains should
also appear naturally in the world volume theories of branes.
Indeed, the BMN matrix model\cite{bmn}, which can be interpretted as
the theory of zero branes in type IIA theory as well as the light
cone supermembrane theory of eleven dimensional supergravity in a
plane wave background\cite{keshav1} is known to be integrable to
rather high orders in perturbation theory\cite{bmn4l}. It provides
us an example of the natural emergence of integrable systems in
brane dynamics. However, a similar systematic approach that utilizes
integrable systems has remained lacking for membrane motions in
other backgrounds. Most notably, membrane motion in trivial/flat
backgrounds have not been studied from the point of view of
integrable models.

To appreciate the chief obstacle that prevents a straightforward
application of quantum spin-chain techniques to matrix regularized
descriptions of membranes in flat spacetimes it is worth recalling
that both the matrix models  mentioned so far, i.e the BMN model and
the dilatation operator of $\mathcal{N}=4$ SYM share a common
feature, which is that they have discrete spectra\cite{bmn,ar1}.
This key common feature is central to being able to analyze them in
the large $N$ limit as quantum spin chains. The discrete nature of
the spectrum of the models mentioned above manifests itself very
transparently at the level of the Hamiltonian in the form of the
presence of quadratic 'mass' terms. One can then go to a basis of
creation/annihilation operators and carry out a perturbative
analysis around the oscillator vacuum. The normal ordered
interaction terms can be understood as Hamiltonians of quantum spin
chains and their contribution to the spectrum can be studied by
applying the Bethe ansatz to the relevant spin chains order by order
in perturbation theory. This connection between quantum spin chains
and large $N$ Hamiltonian matrix models was first worked out by
Rajeev and Lee in\cite{rl}, and it has been extremely useful in the
perturbative analysis of the spectrum and integrability properties
of the BMN matrix model as well as the dilatation operator of
$\mathcal{N}=4$ SYM\cite{keshav1,bmn4l,ar1}. Presence of mass terms
is a feature that however is {\it not} shared by various matrix
models of interest. In the BMN matrix model\cite{bmn} the mass terms
are a reflection of the fact that the background metric for membrane
is a plane wave\cite{keshav1}. Matrix models for membrane motions in
flat spacetimes do not have such quadratic terms. Thus it may seem
that, at least ostensibly, membrane motions in flat spacetimes
cannot be studied by the utilizing the connection of matrix models
to quantum spin chains that has been so successful in the analysis
of the dilatation operator of $\mathcal{N}$=4 SYM and the BMN matrix
model. This is nothing but the standard problem of the existence of
'flat directions' in the Hamiltonian analyses of membrane
motions\cite{hop1,nic1}. However, there are various examples of
membrane Hamiltonians whose spectra are known to be discrete at the
quantum mechanical level even though there are apparent flat
directions at the classical level. The simplest example of such a
scenario is given by the matrix model for the motion of membranes in
a $d+2$ dimensional flat background.\beq H =
\mbox{Tr}\left(\Pi_i\Pi_i -\frac{1}{4} [X_i,X_j]^2\right), i,j =
1\cdots d.\eeq The flat directions correspond to the configuration
of commuting matrices\beq [X_i,X_j]=0\eeq However,  semiclassical
analyses suggest that the spectrum of the model is indeed
discrete\cite{hopth,kl,mem-gap,gab}. In other words, quantum
corrections lift the classical flat directions, and the quantum
effective potential does indeed acquire a mass term. Indeed,
tell-tale signs of integrable behavior in various examples of
membrane motions in flat spacetimes have also been reported in the
past literature\cite{hop2,zac1,zac2,hop3}.  Although, the estimation
of the mass gaps of these models has been carried out using various
techniques in the past, it would be gratifying to have a method that
can enable one to do perturbative analyses around the quantum
corrected effective potentials using the techniques of quantum spin
chains. It is the development of such techniques that we devote the
first part of the paper to. The models that we study to develop our
methods correspond to matrix regularized membrane motions in flat
background spacetimes with and without fluxes. The main summary of
this part of the paper and the corresponding results are as follows.

In section(3) we develop a method for computing the mass gaps in
bosonic matrix models. The method is based on a gauge invariant
rearrangement of various Feynman diagrams of the quantum mechanical
model that allows one to compute the location of the pole of the
propagator order by order in perturbation theory. Similar approaches
for the computation of mass gaps  have also been applied to various
Bosonic matrix models as well as to  supersymmetric models at finite
temperature in the past, (notably in\cite{kl}). One of the key
insights that we gain from our analysis is  the emergence of a new
parameter in the problem which controls the perturbative corrections
to the dynamically generated mass term for the theory as well as the
Hamiltonian perturbation theory around quantum corrected effective
potential for the matrix models. Roughly speaking the parameter
turns out to be $\frac{1}{d}$, where $d$ is the number of matrices
in the problem, which is of course related to  the dimensionality of
the background spacetime for the membrane theory. To put it
differently, the obvious coupling constant of the bosonic matrix
models, i.e the strength of the commutator squared interaction term,
factors out as an overall multiplicative factor in the Hamiltonian
analyses of the theories around the effective potential. However,
one can form a dimensionless number out of the coupling constant of
the matrix model and the dynamical mass term. It is this number that
goes as $\frac{1}{d}$ and appears to control the perturbative
expansion of the matrix models.

As mentioned previously, having a mass term in the effective
potential, as well as a dynamically generated perturbation
parameter, allows one to use the correspondence between large $N$
matrix models and quantum spin chains to compute the large $N$
spectrum of the models in perturbation theory. We carry this out in
section(4). In this part of the paper we  map the  one loop
effective potentials for the relevant models to Hamiltonians of
corresponding quantum spin chains with nearest neighbor
interactions. The spin chains corresponding to membrane motion in
$d+2$ dimensions have $so(d)$ as their invariance group. This is a
reflection of  the $so(d)$ invariance of the interaction
$\sum_{i,j=1}^d \mbox{Tr}[X_i,X_j]^2$ term of the corresponding
matrix models. Having a spin chain emerge as the one loop effective
Hamiltonian of the matrix models is one thing, being able to solve
it and compute the one loop spectrum of the theory is quite another.
If the spin chain turns out to be integrable, in the sense of being
part of a family of mutually commuting Hamiltonians derived from
some underlying $\mathcal{R}$ matrix, then one can use Bethe ansatz
techniques to diagonalize the spin chains\footnote{There are several
books and expository articles on Bethe ansatz techniques. For a
pedagogical introduction, see\cite{fad-rev}}. However, as has been
explained in section(4.1) most of the $so(d)$ invariant spin chains
that emerge as the one loop large $N$ Hamiltonian are {\it not}
integrable in the sense described above. Nonetheless, we can
still make progress using the following two observations. \\
{\bf 1:}For $so(2d)$  or $so(2d+1)$ spin chains with nearest
neighbor interactions, there is always a rather large integrable
sub-sector even if the spin chain is not integrable. This sub-sector
corresponds  the Hilbert space of the spin chain that contains
states charged under $su(d)$ which is contained in $so(2d)$  or
$so(2d+1)$. Since the number of independent excitations of the spin
chains is equal to the rank of the Lie algebra i.e, $d-1$ for
$su(d)$ and $d$ for $so(2d)$, most of the spectrum of the spin
chains of interest to us will be accessible by Bethe ansatz
techniques, even though the spin chains are not strictly integrable.\\
{\bf 2:} If one were interested in the low lying  excitations of the
spin chain, then it is well known that they are captured by a
classical two dimensional non-linear 'sigma' model which can be
thought of as the continuum limit of the spin chain. For $so(d)$
invariant spin chains we can show that the corresponding sigma model
is always integrable even if the quantum spin chain is not. This
understanding is based on realizing the large length $(J)$ limit of
the spin chain as a classical limit and the contraction of the
quantum $\mathcal{R}$ matrices of various $so(d)$ invariant spin
chains to a universal classical $r$ matrix in the large $J$ limit.
Thus one can indeed understand, at least in principle, all the low
lying excitations of the spin chains of interest to us irrespective
of whether or not they are integrable.

The final parts of chapter(4) are devoted to the one-loop
perturbation theory of bosonic models with dynamically generated
mass terms. We identify the integrable sub-sectors of the one loop
spin chains and explicitly diagonalize them using Bethe ansatz
techniques. We also give a formal derivation and proof of the second
statement made above in section(4.4). This is carried out by
deriving the non-relativistic sigma model for the low lying spectrum
of the one loop $so(d)$ invariant spin chains. We establish the
integrability of the sigma models by  deriving the lax pairs and
monodromy matrices for them. This analysis, also brings out, in a
rather explicit form, that the large $J$ limit can be regarded as a
classical limit, with $\frac{1}{J}$ playing the role of $\hbar$,
thus explaining the emergence of classical integrable systems in the
continuum limit.

The final parts of the paper are devoted to the study of
supersymmetric matrix models obtained recently by Kim and Park from
the dimensional reductions of minimally supersymmetric Yang-Mills
theories in diverse dimensions\cite{kp}. The matrix models that we
study in this section have explicit mass terms in them and the focus
of our attention will be the large $N$ integrability of these
models. Indeed, the mechanism for mass generation that we utilize to
study the spectra of various bosonic matrix models will not work in
the case of most supersymmetric matrix models. The mass terms for
the bosonic models arise by summing over certain classes of
self-energy diagrams which will cancel out in the supersymmetric
cases against standard fermionic contributions. However, there does
exist a fairly large class of supersymmetric matrix models with
explicit mass terms in their Hamiltonians; the BMN matrix model is
of course an example of such a scenario. The BMN model is of course
nothing but the light cone M(atrix) theory hamiltonian in the eleven
dimensional plane wave background. It also has at least two other
interpretations. It can also be thought of as the matrix regularized
supermembrane Hamiltonian in the eleven dimensional plane wave
background. As a matrix model, it can just as well also be thought
of as a particular dimensional reduction of minimally supersymmetric
Yang-Mills theory in ten dimensions (or maximally supersymmetric
$\mathcal{N} = 4$ Yang-Mills in D=4) down to one dimension. This
suggests a connection between dimensional reductions of
supersymmetric Yang-Mill theories and super-membnrane models in one
dimension higher than that of the original super Yang-Mills theory.
It is surely instructive to probe this possible connection further.
Minimally supersymmetric Yang-Mills theories can, other than in
$D=10$, be defined in dimensions six, four and three and two, and
the matrix models obtained by the dimensional reduction of these
models to one (time) dimension has recently been accomplished by Kim
and Park\cite{kp}. One has to be a little careful while relating the
dimensional reductions of minimal super Yang-Mills to supermembrane
theories. The naive dimensional reduction of super Yang-Mills to one
dimension is bound to give a supermembrane theory in flat
background. This is of course the original connection made by
BFSS\cite{bfss}. These matrix models do not have mass terms and they
suffer from the problem of flat directions which cannot be cured by
the method we apply to purely bosonic models in this paper. However,
it is possible to add mass terms to the supersymmetric matrix models
in a way that does not break any supersymmetries. The resultant mass
deformed models can then be interpreted as supermembrane theories
{\it not} in flat backgrounds, but rather in plane-wave type spaces.
All possible supersymmetric mass deformations of the matrix models
obtained from the dimensional reductions of minimal super Yang-Mills
theories in dimensions six, four, three and two have been worked out
by Kim and Park in their paper. The existence of the mass terms in
the BMN matrix model is of course what makes it possible to study
its large $N$ spectrum using quantum spin chains. Moreover, the spin
chains obtained in the pertubative expansion of the matrix model are
also known to be integrable to rather high orders in perturbation
theory. It is thus natural to ask what kind of quantum spin chains
arise in the large $N$ perturbation theory for the mass deformed
models obtained in\cite{kp}, and whether or not these spin chains
are integrable. This is the question that we concern ourselves with
in section(5).

Following the analysis carried out by Plefka and
collaborators\cite{bmn4l}we shall focus on the bosonic sectors of
the models obtained by Kim and Park as a starting point for
investigating integrable behavior in the models. In their
paper\cite{kp}, they were able to show that there are two generic
types of possible mass deformations of the matrix models obtained
from the dimensional reductions of minimal Yang-Mills theories. The
mass deformation of the first kind corresponds to matrix models that
have non-trivial maximally supersymmetric vacua. A second type of
mass deformation, which has also been carried out by the same
authors corresponds to supersymmetric matrix models that do not have
non-trivial BPS configurations. In this paper, we are able to show
that all the matrix models that do possess non-trivial
supersymmetric vacua also correspond to integrable spin chains at
one loop. This hints at a strong connection between integrability
and supersymmetry. For the mass deformed models of the second type,
the one loop spin chains are not generically integrable. But, as
mentioned previously, it is always possible to find sectors of the
spin chain that do retain integrability, and thus give us some
information about large $N$ spectrum of the matrix models. This
analysis is carried out in the final section of the paper.
\section{Membranes, Matrix Models and Spin Chains: A Brief Review}
In this section we gather together some known results connecting
membranes to matrix models, and matrix models to quantum spin
chains. A fuller discussion of these connections can be found in
several papers. The relation between membrane motion and Hamiltonian
matrix models is discussed in depth in\cite{hopth, taylor1} while
the use of spin chain techniques in the study of matrix models has
been elaborated upon in\cite{rl,ar1}. We shall refer to these papers
for a more complete account of the results summarized in this part
of the paper.

The action of the membrane in a $d+2$ dimensional flat background is
given by
 \beq S = -T\int d^3\sigma \sqrt{-det h_{\alpha \beta
}},\eeq where $T$ is the brane tension, which can be expressed in
terms of the basic dimensional parameter in the problem, the Planck
length, $l_p$ as  \beq T = \frac{1}{2\pi l_p^3}.\eeq $h_{\alpha
\beta } =
\partial _\alpha X^\mu \partial _\beta X_\mu $ is the pull back of
the spacetime metric to the membrane world volume. $\mu, nu$ take on
values $1\cdots d+2$ while $\alpha, \beta $ are the three world
volume indices. The Nambu-Goto type membrane action can be replaced
by a Polyakov action at the expense of the introduction of a
fiducial world volume metric $\gamma $, and the action can be
written as\beq S = \frac{T}{2}\int d^3\sigma \sqrt{-\gamma
}\left(\gamma ^{\alpha \beta }\partial _\alpha X^\mu
\partial _\beta X_\mu -1\right).\eeq The three diffeomorphism
symmetries corresponding to the three worldvolume coordinates can be
used to eliminate three of the six metric components, which are
fixed as\beq \gamma_{0i} = 0, \gamma _{00} = -\frac{4}{\nu
^2}\mbox{det}h.\eeq One can introduce light front coordinates  \beq
X^{\pm } = \frac{1}{\sqrt{2}}(X^1 \pm X^{D+2}),\eeq and choose the
light front gauge $X^+ = \tau$ which eliminates two of the $d+2$
fields. The gauge fixed Hamiltonian in the light front gauge takes
on the form\beq H = \frac{\nu T}{4}\int d^2\sigma \left(\dot{X}
^i\dot{X}^i + \frac{2}{\nu ^2}\{X^i,X^j\}\{X^i,X^j\}\right).\eeq In
the above formula, $X^i$ are the transverse degrees of freedom and
there are $d$ of them. The curly brackets are the familiar Poisson
brackets on the membrane surface, signifying the invariance of the
action under the algebra of area preserving diffeomorphisms. The
gauge fixed Hamiltonian can be regularized by replacing the algebra
of area preserving diffeomorphisms on the membrane surface by that
of finite $N\times N$ matrices\cite{hopth}. For specificity, one can
take the membrane to have the topology of $S^2\times R$ in which
case one can approximate the algebra of area preserving
diffeomorphisms on the two sphere by matrices which transform in the
$N$ dimensional irreducible representation of $su(2)$. The free
parameter $\nu $ which was introduced in the problem while gauge
fixing can be set to $N$. The details of the the approximation
method leading to the regularization of membrane motions by the
quantum mechanics of matrices can be found at various places in the
literature; see for example\cite{hopth,taylor1}. The final answer
for the regularized form of the membrane Hamiltonian in $D+2$
dimensions is given by the quantum mechanics of $D$ hermitian
matrices as\beq H = g^3 \mbox{Tr} \left(\Pi _i \Pi_i \right) -
\frac{1}{4g^3}\mbox{Tr}\left([X^i,X^j][X^i,X^j]\right)\eeq where\beq
g^3 = 2\pi l_p^3.\label{flmem}\eeq

One could also allow the background spacetime to have fluxes to
which the membrane degrees of freedom can couple. To take the
simplest  concrete example, one could look at zero-brane quantum
mechanics of type IIA string theory in the presence of a non
vanishing vev of the four form flux.  The dynamical model is
described by the quantum mechanics of three hermitian matrices
$X_i$\cite{my}, the Euclidean action for which is \beq S = \int dt
TTr\left(\frac{1}{2}\dot{X}_i^2 - \frac{1}{4}[X_i,X_j]^2 +
\frac{i\kappa}{3}\epsilon_{ijk}X_iX_jX_k\right).\eeq The cubic
interaction term expresses the interaction between the matrix model
and the four form flux. The constant four form flux is taken to
be\beq F^{(4)}_{tijk} = -2\kappa \epsilon_{ijk}.\eeq $T =
\frac{\sqrt{2\pi}}{g_s}$ is the zero brane tension. It is convenient
to scale the matrices $X_i \rightarrow \frac{1}{\sqrt{T}}X_i$ and
define\beq \frac{1}{\sqrt{T}} = g\eeq so that the Euclidean space
action takes on the form  \beq S = \int dt
Tr\left(\frac{1}{2}\dot{X}_i^2 - \frac{g^2}{4}[X_i,X_j]^2 + \frac{ig
\kappa}{3}\epsilon_{ijk}X_iX_jX_k\right).\label{action}\eeq In the
action the dimension of the matrices $X_i$ is $\sqrt{L}$ while those
of $g^2$ and $\kappa $ are $\frac{1}{L^3}$ and $\frac{1}{L}$
respectively. Matrix models with similar cubic couplings also arise
when one considers M(atrix) theories in plane wave backgrounds i.e
the BMN matrix model and in mass-deformed matrix models obtained
from the dimensional reduction of super Yang-Mills theories. For the
purposes of this paper, we could turn on such a four-form flux and
couple it to the motion of a membrane in a $d+5$ dimensional flat
background. The flux, will pick out three special directions, and at
the level of matrix quantum mechanics, three of the $d+3$ matrices,
will couple to the background flux. The resultant matrix mechanics
will be a straightforward generalization of the Myers' model
described above. The resultant action would be  \beq S = \int dt
Tr\left(\frac{1}{2}\dot{X}_i^2 -
\sum_{i,j=1}^{d+3}\frac{g^2}{4}[X_i,X_j]^2 + \frac{ig
\kappa}{3}\sum_{a,b,c
=1}^3\epsilon_{abc}X_iX_jX_k\right).\label{flmemfl}\eeq Several
other examples of matrix mechanics involving cubic Chern-Simons type
couplings will also be studied in the final parts of the paper
in the context of supersymmetric matrix models.\\
{\bf From Matrix Models to Spin Chains:}\\
Let us now briefly review the connection between the perturbative
dynamics of Hamiltonian matrix models and quantum spin chains. This
connection, which was first established by Rajeev and Lee\cite{rl},
 will be central to the computation of the spectrum of various
models of interest to us. Let us consider a matrix model Hamiltonian
of the kind\beq H = \mbox{Tr}\left(\frac{1}{2}(\Pi_i^2 +
\mu^2\Phi_i^2) + \frac{1}{N}\Psi ^4_{ijkl}\Phi ^i\Phi^j\Phi^k\Phi^l
+ \frac{1}{\sqrt{N}}\mu \Psi ^3_{ijk}\Phi ^i\Phi^j\Phi^k\right)\eeq
The tensors $\Psi ^3,\Psi^4$ encode information about the cubic and
quartic interaction terms. Looking ahead at the prospect of taking
the 't Hooft large $N$ limit in various analyses that we shall
perform, we have incorporated various factors of $\frac{1}{N}$ in
the generic Hamiltonian above so that it does posess a well defined
't Hooft large $N$ limit. The chief qualitative difference in the
nature of the Hamiltonian given above and the two examples discussed
previously (\ref{flmem},\ref{flmemfl}) is the quadratic 'mass' term.
The dynamical emergence of such mass terms is of course what a
substantial part of the paper will be devoted to. Keeping later
applications in mind we recall how the perturbative large $N$
analysis of matrix model Hamiltonians becomes a problem of
diagonalizing quantum spin chains.

Let us start by introducing  the matrix creation and annihilation
operators \beq A_i = \frac{1}{\sqrt{2}\mu}\left(\mu X_i +
i\Pi_i\right)\eeq and their adjoints, in terms of which  the
Hamiltonian takes on the form \beqs H = \mbox{Tr}\left(\mu
A^{\dagger i}A_i + \frac{1}{4N\mu ^2}\Psi^4_{ijkl}[A_i + A^{\dagger
i}][A_j + A^{\dagger j}][A_k +
A^{\dagger k}][A_l + A^{\dagger l}]\right)\nonumber\\
+\left(\Psi^3_{ijk}\frac{1}{(2^{3/2}N^{1/2}\mu ^{1/2})}(A_i +
A^{\dagger i})(A_j + A^{\dagger j})(A_k + A^{\dagger
k})\right).\label{normal}\eeqs We shall be interested in computing
the first order correction to energies of single trace states such
as\beq |i_1i_2\cdots i_J> =
\frac{1}{N^{J/2}}\mbox{Tr}\left(A^{\dagger i_1}A^{\dagger i_2}\cdots
A^{\dagger i_J}\right)|0>.\eeq All such single trace states are
eigenstates of the free Hamiltonian with an eigenvalue
$\mathcal{E}_0 = \mu J$. The perturbative corrections to the
energies of the free Hamiltonian can be arranged in an expansion in
powers of $\frac{1}{\mu ^2}$. \beq \mathcal{E} = \mu J +
\frac{1}{\mu ^2}\mathcal{E}_1 + \cdots \eeq For an eigenstate of the
free Hamiltonian $|\mathcal{I}>$, \beq \mathcal{E}_1 =
<\mathcal{I}|V|\mathcal{I}> \mbox{where } V = V^4+ V^3.\eeq  $V$ is
the vertex that contributes to the first order energy shift, and it
can be built out of the cubic and quartic terms in (\ref{normal})
as\beq V^4 = H^4, V^3 = \frac{H^3(I-\Pi)H^3}{H_0 - J}. \eeq $\Pi$ is
the projector to the subspace of the Hilbert space orthogonal to the
one spanned by states of length $J$\cite{bmn4l}, while $H^3, H^3$
are the quartic and cubic terms in (\ref{normal}) respectively.

Since only the diagonal matrix elements of$V$ matter for first order
perturbation theory, it is sufficient to keep those terms in $H^4$
and $H^3$ that have an equal number of creation and annihilation
operators. Not all such terms are important if the large $N$ limit
is taken.  The terms in $V$ that have leading order matrix elements
of $O(1)$ are the ones for which normal ordering is compatible with
the ordering implied by the trace. i.e they are terms that involve a
string of creation operators followed by a string of annihilation
operators sitting inside a trace. We can denote such operators by
the symbol $\Theta $; for example,\beq \Theta ^{i,j}_{k,l} =
\frac{1}{N}Tr\left(A^{\dagger i}A^{\dagger j}{A_lA_k}\right).\eeq
The action of these operators on single trace states can be written
down in closed form.\beq \Theta ^{i,j}_{k,l}|i_1\cdots i_J> =
\sum_{m=1}^J\delta ^{i_m}_k\delta ^{i_{m+1}}_l|i_1\cdots
i_{m-1}iji_{m+2}\cdots i_J>+O(\frac{1}{N}).\eeq In other words the
operators $\Theta $ can be thought of as a 'machine' that runs along
the entire length of the state and checks if any neighboring indices
match the lower indices of the operators, and if they do, it
replaces the neighboring indices by the upper indices. This leads to
a map between large $N$ matrix models and quantum spin chains. The
single trace states can be thought of as states of a quantum spin
chain with periodic boundary conditions, the periodicity being
inherited from the cyclicity of the trace. The indices $i_1\cdots
i_J$ are to be thought of as the 'spins' of the spin chain, which
take on two values. A general map between quantum spin chains and
large $N$ matrix models has been proposed and discussed at length in
\cite{rl} and it has been also used in some detail to study
integrability of the dilatation operator of $\cal{N}$ $=4$ super
Yang-Mills theory in\cite{ar1,ar2}; however for the purposes of the
present discussion it is sufficient to restrict ourselves to spin
chains with nearest neighbor interactions. The $so(d)$ invariant
operators obtained from $V$, that survive the large $N$ limit are $
\Theta ^{i,j}_{i,j}, \Theta ^{i,j}_{j,i}$ and$ \Theta ^{i,i}_{j,j}$.
It is clear that $\Theta ^{i,j}_{i,j}$ acts as the number operator,
thus in the large $N$ limit\beq \Theta ^{i,j}_{i,j} = Tr(A^{\dagger
i}A_i) = J\eeq The other two operators can be mapped to specific
spin chain operators. The action of these spin chain operators on
neighboring spins is\beqs \Theta ^{i,j}_{j,i} &=& P_{l,l+1};
P(i\otimes j) = j\otimes i\nonumber \\
\Theta ^{i,i}_{j,j} &=& K_{l,l+1}; K(i\otimes j) =
\delta_{i,j}\sum_m(m\otimes m).\eeqs In the above equations the
arguments of $K$ and $P$ are to be interpreted as spins sitting on
neighboring sites. $P$ permutes neighboring spins and $K$ traces
over spin values . The large $N$ spectrum at one loop for a general
$so(d)$ invariant matrix model is given by a nearest neighbor spin
chain of the following generic form. \beq H_{1-loop} = \mu I_{l,l+1}
+ \frac{1}{\mu ^2}\sum_l\left(\alpha I_{l,l+1} + \beta P_{l,l+1} +
\gamma K_{l,l+1}\right)\eeq The coefficients $\alpha, \beta ,\gamma
$ carry information about the specific details about the matrix
model being studied. To avoid various divergences that might
potentially arise due to normal ordering issues, it is useful to
compute {\it not} the absolute energies of various states , but the
energies of various states relative to a reference state. Since
$\sum_l I_{l,l+1} = J$ for all states of a given length, all terms
proportional to the identity operator drop out when one measures the
energies with respect to a particular state. The one loop spin
chains that accomplish the task of measuring the energies with
respect to a reference state are of the form\beq \Delta ^1 =
\frac{1}{\mu ^2}\sum_l\left( \beta P_{l,l+1} + \gamma
K_{l,l+1}\right)\eeq

Of great interest in membrane dynamics are the following type of
couplings. \beq \Psi ^4_{ijkl} =
-\frac{1}{2}\left(\delta_{ik}\delta_{jl}-\delta_{il}\delta_{jk}\right)\eeq
so that the quartic term can be written as
$-\frac{1}{4}[\Phi^i,\Phi^j]^2$. The particular cubic coupling that
will be of importance later in the paper is of the Chern-Simons
type\beq \Psi^3_{ijk} = i\nu\epsilon_{ijk},\eeq where $\nu $ is a
numerical constant. Keeping in mind the various models that we are
going to study, we shall list $\Delta ^1$ for these two  classes
of models\\
{\bf I:} $so(d)$ invariant matrix models for which $\Psi ^3 =0$.
\beq \Delta ^1 = \frac{1}{2\mu ^2}\sum_l \left( \frac{1}{2}K_{l,l+1}
- P_{l,l+1}\right).\label{ensod}\eeq {\bf II:} $so(3)$ invariant
models for which $\Psi ^3_{ijk} = -i\nu \epsilon_{ijk}$.\beq \Delta
^1 = \frac{1}{2\mu ^2}\sum_l \left((3\nu ^2-1)P_{l,l+1} +
\frac{1}{2}(1-9\nu ^2)K_{l,l+1}\right).\label{enso3}\eeq

\section{Dynamical Mass Generation via Resummed Perturbation Theory}
We shall now revert back to the original matrix models for flat
space membrane dynamics (\ref{flmem},\ref{flmemfl}) and bring them
to a form such that their spectra can be analyzed by quantum spin
chain techniques. The crucial ingredient for this to be possible is
the presence of a mass term in the matrix model Hamiltonian, the
dynamical generation of which shall be the focus of this section. As
outlined previously, we would like to set up a well defined
perturbation theory for estimating the mass-gaps as well as the
spectra of the matrix models of interest to us. On the face of it
this might appear to be a self-contradictory goal. Dynamical mass
generation is usually regarded as a non-perturbative phenomenon.
Indeed, the methods that have been employed to estimate mass-gaps in
matrix models in the past, see for example\cite{kl}, do indeed make
use of non-pertrubative techniques, such as Schwinger-Dyson
equations, to estimate the spectral gaps. On the other hand, the
spin-chain techniques, that we ultimately want to make use of to do
a large $N$ Hamiltonian analysis  appear to be an efficient method
of doing the usual perturbative expansion. A way out of this impasse
is provided is provided if one can discover a new 'hidden'
parameter, which in our case will turn out to be $\frac{1}{d-1}$, in
which the spectrum of excitations as well as the mass gap of the
theory can be expanded. Indeed, one of the key points that we would
like to bring out from this analysis is the emergence of this new
parameter and show that it  controls the perturbation theories
employed both to estimate the mass gap as well as the one used for
carrying out a Hamiltonian quantization of the theory around the
quantum corrected effective potential.

We start with  the Euclidean action for membrane motion without
fluxes(\ref{flmem}), which is\beq S = \int dt
\frac{1}{g^3}Tr\left(\frac{1}{2}\dot{X}_i\dot{X}_i -
\frac{1}{2N}\sum_{j>i}[X_i,X_j]^2\right).\label{eucs} i,j = 1\cdots
d-2.\eeq We now add and subtract a mass term to the Euclidean action
\beq S = S_m - \hbar \int dt \frac{1}{2g^3}m^2Tr (X_iX_i)\eeq where
\beq S_m = S +\frac{1}{g^3}\int dt \frac{1}{2}m^2Tr (X_iX_i).\eeq
The goal is not to change the theory but only to rearrange the
Feynman diagrams. Thus $\hbar $ will have to be set equal to one. We
shall use the numerical value of $\hbar $ only at the end of the
computation. Its role is that of a loop counting parameter to help
us
carry out an efficient reorganization of the Feynman diagrams.\\
The action, after appropriate insertions of factors of $\hbar $
reads as\beq S = \int dt\frac{1}{2g^3}Tr\left(\dot{X}_i\dot{X}_i +
m^2X_iX_i- \frac{\hbar}{2N} [X_i,X_j]^2\right) - \hbar \int dt
\frac{ m^2}{2g^3} Tr(X_iX_i).\eeq Clearly, this is nothing but the
original action if $\hbar $ is set to one. We shall let  $m^2$ admit
an expansion in $\hbar $. \beq m^2 = m_1^2 + \hbar m_2^2 + \hbar
^2m_3^2 + \cdots \eeq The propagator can be read off and it is\beq
G_2(p) = \left<(X_i)^a_b(p)(X_j)^c_d(-p)\right> = \frac{\delta
^a_d\delta ^c_b\delta _{ij}}{\frac{1 }{g^3}(p^2 + m^2 - \hbar m^2) +
\Sigma(p)},\eeq where $\Sigma (p)$ is the self-energy correction to
the propagator. Clearly, $\Sigma (p)$ has an expansion in powers of
$\hbar$, \beq \Sigma (p) = \hbar \Sigma _1(p) + \hbar ^2 \Sigma
_2(p) + \cdots \eeq We can now invoke a self-consistency argument,
which is that if $m$ is indeed the gap in the spectrum i.e the
location of the pole of the propagator then looking at the form of
the propagator above, we see that $m_1, m_2$ etc must chosen to
cancel the perturbative correction to the propagator that arise from
the self energy diagrams $\Sigma _1, \Sigma _2\cdots $ order by
order in perturbation theory. This sets up a series of gap equations
that determine the mass gap of the matrix model\footnote{This
generic technique is known to work rather well for gauge theories as
well. For example it has been used to get a remarkably good estimate
of the mass gap of pure Yang-Mill in 2+1 dimensions in\cite{an}.}.

So far, we have set up the problem of determining the mass gap as a
perturbative computation in $\hbar$, but of course, $\hbar $ is an
invented parameter and its numerical value is one.   That $\hbar$ is
a reasonable parameter to do perturbative  computations in will have
to be validated by an explicit demonstration that the corrections to
the mass, when thought of as an expansion in this artificial
parameter, turn out to be small in comparison to the leading order
or one-loop estimate. That is indeed the case, and we explicitly
demonstrate that in the next section. As a matter of fact, a two
loop computation which is reported below, suggests that what appears
to be an expansion in $\hbar$ is really an expansion in
$\frac{\hbar}{d-1}$. This is indeed gratifying, as this generates
for us a bona-fide parameter in which to carry out perturbative
computations of the mass gap in the problem. We shall now proceed to
carry out explicit perturbative computations for the mass gap and
arrange the results in a $\frac{1}{d-1}$ expansion. In what follows,
we shall only focus on the planar diagrams, which give the leading
order results for the mass-gap in $\frac{1}{N}$.

\subsection{Perturbative Corrections to the Mass:}
{\bf One Loop Mass-Gap:}\\The one loop correction to the mass is
given by the condition that $m_1^2$ cancel out the one-loop self
energy diagram:\beq \frac{\hbar  m_1^2}{g^3} = \hbar
\Sigma_1.\label{1lpgap}\eeq $\Sigma_1$ is given by the standard
'tadpole' integral\beq \Sigma_1 = \Upsilon_1\int \frac{dp}{2\pi}
\frac{1}{p^2 + m_1^2} \eeq $\Upsilon_1$ is the standard
combinatorial factor that counts the number of diagrams contributing
to the tadpole graph in the large $N$ limit. This number may be
computed easily enough. To do that, we first observe that out of the
two terms in the potential energy, $Tr(X_iX_jX_iX_j)$ and
$Tr(X_iX_iX_jX_j)$, only the the second one contributes in the large
$N$ limit. That is the so, because, it is only in this diagram that
the internal momentum loop of the tadpole diagram results from the
Wick contraction of two neighboring $X$ fields sitting inside a
trace, which  leads to a contribution of order $N$. Let us suppose,
we were computing the correction to the $<X_1X_1>$ propagator, then
clearly, the number of ways in which a $X_1$ field from the
$Tr(X_1X_1X_jX_j)$vertex can be attached to an external line
corresponding to the propagator is two. There is also an overall
factor of $d-1$ from the $X$ fields running in the loop. Thus \beq
\Upsilon_1 =2(d-1).\eeq Thus the gap equation at the one loop order
gives the value of the mass-gap to be\beq m_1^3 =
(d-1)g^3.\label{1mass}\eeq {\bf Two
Loop Correction to The Mass:}\\
At the two loop level, there are two different sources of
contributions to terms of $O(\hbar ^2)$. Since $\hbar $ appears
explicitly in the formula for the propagator, the tadpole diagram,
which goes as \beq \int\frac{dp}{2\pi} \frac{1}{p^2 + m^2-\hbar
m^2}\label{tad}\eeq will generate a term of order $\hbar ^2$. Apart
from this there is the usual two loop contribution due to the
'double-scoop' and 'sunset' diagrams\footnote{We are borrowing the
terminology of\cite{ramb} for the Feynman diagrams}. Thus \beq
\Sigma_2 = \frac{1}{2} \frac{(d-1)}{m} - \frac{g^3}{
m^4}I(p)\label{sigma2}\eeq The first term in the expression for
$\Sigma_2$ above comes from the expansion of the formula for the
'one loop' tadpole diagram  in higher powers of $\hbar $, while the
second term arises from the usual double scoop and sunset two loop
Feynman graphs. $I(p)$ being the total contributions of the Feynman
integrals, discussed in the appendix. It suffices for now to know
the numerical value of these diagrams, which is, \beq I(p)
=\frac{1}{2}((d-1)^2+(d-1)). \eeq (\ref{sigma2}) explicitly,
demonstrates the intertwining of terms, that would naively be
considered 'one-loop' diagrams, such as the 'tadpole' diagram with
two loop diagrams in the second order contribution to the correction
of the mass. Thus the two loop gap equation is:\beq \frac{\hbar
(d-1)}{m} + \frac{\hbar^2(d-1)}{2m} - \frac{\hbar^2g^3}{2
m^4}((d-1)^2+(d-1)) = \frac{\hbar m^2}{g^3}\label{2lpgap}\eeq
  To solve this equation perturbatively, we
can set\beq m^3 = g^3(d-1)\left( 1 + \hbar \delta\right).\eeq The
solution to $O(\hbar )$ is \beq \delta = -\frac{1}{2(d-1)}\eeq Thus
the two loop correction to the mass can be written as\beq m =
((d-1))^{1/3}g(1- \frac{\hbar}{6(d-1)} +\cdots) = m_1(1-
\frac{\hbar}{6(d-1)} +\cdots).\label{mass}\eeq Hence, we see that
the two loop correction to the mass is explicitly suppressed by a
factor of $\frac{1}{d-1}$. Thus, as advertised before, the
computation of the mass gap can indeed be organized in a
$\frac{1}{d-1}$ expansion, at least to the two loop order.

Even when $\frac{1}{d-1} =1$, we see that the combinatorics and the
numerical values of the Feynman integrals involved add up to
suppress the two loop contribution. As argued in \cite{kl} it is
possibly too much to expect that the perturbation series for $m$
actually converges, however it can possibly be regarded as a good
asymptotic expansion, of which, we shall keep only the leading order term in later spin-chain computations.\\
{\bf Aside on the Two Loop Feynman Integrals:}\\
The contributions of the 'double scoop'  and the 'sunset' diagrams
can be written as\beq \frac{g^3}{ m^4}(\Upsilon _2I_2 +
\Upsilon_3I_3)\eeq where $\Upsilon_{2,3}$ are the combinatorial
factors that count the number of diagrams contributing to the
'double scoop' and the 'sunset ' graphs at large $N$. $I_{2,3}$ are
the relevant Feynman integrals\beqs I_2 =
\int\frac{dpdp'}{(2\pi)^2}\frac{1}{(p^2+1)(p^{'2}+1)^2}=\frac{1}{8}\eeqs
and\beqs I_3
=\int\frac{dpdp'}{(2\pi)^2}\frac{1}{(p^2+1)(p'^2+1)((p-p')^2 +1)} =
\frac{1}{4\pi^2}\int_{-\pi/2}^{\pi/2}dx\int_{-\pi/2}^{\pi/2}dx'\frac{1}{(\tan(x)-\tan(x'))^2
+1} = \frac{1}{12}\eeqs $\Upsilon_2$ is the number of Feynman graphs
contributing to the double scoop diagram. This can be computed as
follows. Once again, let us assume that we are computing the
correction to the $<X_1X_1>$ propagator. Once again it is easy to
convince oneself that the only vertex contributing to the double
scoop diagram, just as in the case of the tadpole diagram is
$Tr(X_1X_1X_jX_j)$. The number of ways in which one of the $X_1$
fields can be contracted with a an external leg of the propagator is
two. The number of ways the two $X_j$ fields can be contracted with
the corresponding fields of the second vertex is also 2. One also
picks up a factor of $d-1$ from each loop. Thus\beq \Upsilon_2
=4(d-1)^2.\eeq In the case of the sunset diagram, we shall have to
consider the cases when both the vertices are of the
$Tr(X_iX_iX_jX_j)$ and that when both are of the $Tr(X_iX_jX_iX_j)$
type. Counting along similar lines as the case above shows that the
total number of planar diagrams contributed by the first instance is
$2(d-1)$ and the number when both the vertices are of the
$Tr(X_iX_jX_iX_j)$ type is $4(d-1)$. The case that involves mutual
contractions between the vertices of these two types is of lower
order in $\frac{1}{N}$. Thus \beq\Upsilon_3 = 6(d-1),\eeq leading to
\beq \Upsilon_2I_2+\Upsilon_3I_3 =\frac{1}{2}(d-1)\eeq which was
used in (\ref{2lpgap}).

\subsection{Mass-Gaps For Models With Chern-Simons Couplings:} The above analysis can
be easily extended to estimate the mass gap in flat space membrane
motion in the presence of fluxes i.e to the generalization of the
Meyers' model described in(\ref{flmemfl}). The action that we now
consider is \beq S = \int dt
Tr\left(\frac{1}{2}\dot{X}_i^2-\frac{g^2}{4}[X_i,X_j]^2 + \frac{ig
\kappa}{3}\epsilon_{abc}X_aX_bX_c\right).\eeq In the commutator
squared interaction term of the action, $i,j = 1\cdots d+3$ while
for the Chern-Simons coupling $a,b,c = 1,2,3$. The total number of
Bosonic matrices in the problem is $d+3$, $d$ of which transform
under $so(d)$, while three of the matrices which are picked out by
the anisotropy induced by the flux transform under $so(3)$. The
symmetry group of the problem is obviously $so(d)\times so(3)$. The
one-loop mass for the $so(d)$ scalars is the same as before.
Adapting the result(\ref{1mass}) to the present problem we can read
off the one loop masses of the $so(d)$ scalars to be\beq m^3_{so(d)}
= Ng^2(d+2).\eeq To do a one-loop estimation of the mass gap for the
$so(3)$ scalars we shall have two Feynman diagrams to consider. The
first one corresponds to one insertion of the quartic vertex and the
second one  involves insersions of the cubic vertex. The leading
large $N$ contribution of the insertion of the quartic vertex has
already been discussed at length in the previous section.
 As far as the cubic vertices are concerned, it can be seen, upon
carrying out the contraction of the color indices that the leading
order large $N$ contribution comes from the
$Tr(X_1X_2X_3)Tr(X_1X_3X_2)$ vertex. The momentum integrals are all
convergent and can be evaluated in closed form by elementary means.
The resulting gap equation reads as\beq (d+2)\lambda + \frac{\lambda
\kappa ^2}{m_{so(3)}^2} = m_{so(3)}^3,\eeq where, \beq \lambda =
Ng^2\eeq is the 't Hooft coupling of the matrix model. To solve this
equation we first observe that $\kappa ^2 $ and $\lambda ^{2/3}$
have the same dimensions. Thus, we can define a dimensionless
parameter $\beta $ such that \beq \kappa ^2  = \beta \lambda
^{2/3}.\eeq In the regime where the strength of the four form flux
is weak, $\beta < 1$, we can solve the gap equation order by the
ansatz\beq m_{so(3)}^3 = C\lambda f(\beta ), f(\beta ) = 1+f_1\beta
+ f_2\beta^2 +\cdots\label{mass3}\eeq The gap equation now becomes a
recursion relation for the coefficients $f_i$ while $C=d+2$. The
first few coefficients are\beq f_1 = \frac{1}{(d+2)^{5/3}}, f_2 =
\frac{2}{3((d+2)^{5/3})}f_1.\eeq The higher coefficients in the
series fall off as powers of $\frac{1}{C^{5/3}}$. Thus the mass gaps
for the $so(3)$ scalars arranges itself in a $\frac{1}{d}$
expansion, and to leading order in $\frac{1}{d}$ all the $d+3$
scalars of the problem have the same mass.
\section{Hamiltonian Analysis of the Matrix Models} Having estimated
the mass gaps for various Bosonic matrix models, we are now in a
position to carry out a Hamiltonian quantization of these models and
study their large $N$ spectrum using the techniques of quantum spin
chains. Let us begin by deriving the one-loop energy operator
$\Delta_1$ for the most general Bosonic model analyzed so far, which
is the $so(d)\times so(3)$ symmetric model considered in the
previous section.

After incorporating the dynamically generated mass term
(\ref{mass3}), the Hamiltonian of the  matrix model model
becomes\beq H = Tr\left(\frac{1}{2}\dot{\Pi}_i^2
+\frac{m^2}{2}X_iX_i-\frac{g^2}{4}\sum_{i,j=1}^{d+3}[X_i,X_j]^2 +
\frac{ig \kappa}{3}\sum_{a,b,c
=1}^3\epsilon_{abc}X_aX_bX_c\right).\eeq In writing down the above
Hamiltonian, it is of course implied that we are starting from an
effective action in which the self energy corrections to the
propagators have been accounted for at the one-loop level. Since,
the method of re-summed perturbation theory does not affect the
vertices of the matrix model, they remain unchanged at the level of
the Hamiltonian as well. The only modification is the inclusion of
the dynamically generated mass term, which is the same (\ref{mass3})
for all the $d+3$ scalars at the one loop level.

The mass terms completely lift the classical flat directions and one
can now proceed to  compute the energy operator.  As outlined in the
'introduction' the perturbative computation is best done in the
basis of   matrix creation and annihilation operators \beq A_i =
\frac{1}{\sqrt{2}m}\left(mX_i + i\Pi_i\right), A^{\dagger}_i=
\frac{1}{\sqrt{2}m}\left(mX_i - i\Pi_i\right).\eeq The Hamiltonian
becomes\beq H = \lambda ^{1/3}\mbox{Tr}\left(\mu A^{\dagger i}A_i -
\frac{1}{16N\mu ^2}[A_i + A^{\dagger i},A_j + A^{\dagger j}]^2 +
i\frac{\sqrt{\beta }}{3(N^{1/2}2^{3/2}\mu ^{3/2})}\epsilon_{ijk}(A_i
+ A^{\dagger i})(A_j + A^{\dagger j})(A_k + A^{\dagger
k})\right)\eeq where $\mu $ is defined to be \beq m = \lambda
^{1/3}\mu =  \lambda ^{1/3}(Cf(\beta ))^{1/3}.\eeq Since the
symmetry of the problem splits into $so(d)\times so(3)$ it makes
sense to write down the energy operator for the $so(3)$ and the
$so(d)$ part of the Hamiltonian separately. That can be done easily
enough using the basic formulae (\ref{ensod},\ref{enso3}); and the
result is\beq \Delta_{1,so(d)} = -\frac{1}{2\mu
^2}\sum_l\left(P_{l,l+1}-\frac{1}{2}
K_{l,l+1}\right).\label{bose1}\eeq \beq \Delta_{1,so(3)}
=\frac{1}{2\mu ^2}\sum_l\left((3\theta^2 -1)P_{l,l+1} +
\frac{1}{2}(1-9\theta^2)K_{l,l+1}\right), \theta^2 =
\frac{\beta}{\mu^{2}}.\label{bose2}\eeq It is worth noting that $\mu
$ is essentially the same quantity that controls the perturbative
expansion of the mass-gap discussed earlier. From the above
formulae, it is also clear that $\mu $ is the dimensionless
parameter that controls the Hamiltonian perturbation theory as well.
\subsection{Integrability:} Having a spin-chain formalism for the
one-loop energy operator does not automatically imply that we shall
be able to solve for its spectrum. The obvious obstacle in this
direction is that generic $so(d)$ invariant quantum spin chains are
{\it not} integrable by the methods of the algebraic Bethe ansatz.
For an $so(d)$ invariant quantum spin chain, with nearest neighbor
interactions, $d>2$ and spins in the defining representation, to be
integrable, the Hamiltonian has to be of the following
form\cite{mz,resh1}\beq H = \alpha \sum_l\left(g'P_{l,l+1} +
K_{l,l+1}\right), g' = (1-\frac{d}{2}).\eeq $\alpha $ is an
arbitrary coupling constant. One is of course free to add a term
proportional to the identity operator to the Hamiltonian and not
loose integrability. It is easy to see that neither one of
({\ref{bose1},{\ref{bose2}) is of the form required by
integrability. However, one can still make progress, as
non-integrable spin chains also contain sectors  can indeed be
studied by the Bethe ansatz. To proceed further, let us briefly
recall the construction of integrable spin chains starting from an
underlying $\mathcal{R}$ matrix, which for $so(d)$ spin chains
is\cite{mz,resh1}\beq \mathcal{R}_{12}(u) = u(\hbar g' - u)I_{12} +
\hbar(\hbar g' - u)P_{12} + \hbar u K_{12}.\eeq $g' =
(1-\frac{d}{2})$. For later uses, we have put in a parameter $\hbar$
in the $\mathcal{R}$ matrix to act as a bookkeeping device. The
numerical value of $\hbar $ is one and it is nothing but the
strength of non-commutativity of the spin operators  at the same
lattice point.  The $\mathcal{R}$ matrix is an operator that acts on
a tensor product of two vector spaces both of which are
$\mathcal{C}^d$ when the spins are in the defining representation of
$so(d)$. The indices $1,2$ refer to the respective vector spaces,
while $u$ is the spectral parameter\footnote{For a pedagogical
introduction to the use of $\mathcal{R}$ matrices in the
construction of integrable spin chains, see\cite{fad-rev}.}. For a
spin chain with $J$ sites the transfer matrix is built out of the
$\mathcal{R}$ matrix as\beq \mathcal{T}(u)
=\mathcal{R}_{01}(u)\mathcal{R}_{02}(u)\cdots
\mathcal{R}_{0J}(u).\eeq The subscript $0$ refers to the auxiliary
space, which is not a part of the physical Hilbert space of the spin
chain, and it will be traced over. Let $Tr_0$ denote the trace over
the auxiliary space. The transfer matrix can be obviously be
expanded in powers of the spectral parameter \beq \mathcal{T}(u) =
\sum_m u^m\mathcal{T}_m\eeq and taking the trace over the auxiliary
space of both sides of the above equations produces a generating
function for the local conserved charges of the spin chain. \beq
t(u) = Tr_0\mathcal{T}(u) = \sum_m u^m t_m.\eeq Finding a closed
form of for the local charges is usually a tedious but
straightforward computation;  however, we shall be interested in
only the first few charges, especially the second one which
corresponds to the Hamiltonian. The first charge is\beq t_0 =
Tr_0\prod_{l=1}^J(\hbar ^2g')^JP_{0l}\eeq It can be shown that this
charge corresponds to the discrete lattice shift operator.\beq
t_0(\mathcal{V}_1\otimes \mathcal{V}_2\otimes \cdots \otimes
\mathcal{V}_J) =(\hbar ^2g')^J(\mathcal{V}_2\otimes
\mathcal{V}_3\otimes \cdots\mathcal{V}_J\otimes \mathcal{V}_1).\eeq
The next term in the expansion is \beq t_1 = \sum_{l=1}^J
t_0\left(\hbar g' P_{l,l+1} - \hbar I_{l,l+1} + \hbar
K_{l,l+1}\right).\eeq Since $t_0$ is a constant of motion and we are
free to add constant terms to the Hamiltonian it is easy to see that
the most general form for the nearest neighbor Hamiltonian with a
$so(n)$ symmetry is\beq H = \alpha \sum_{l=1}^J\left( g'P_{l,l+1}
-\beta I_{l,l+1} + K_{l,l+1}\right) \label{intsod}\eeq where $\alpha
$ and $\beta $ are arbitrary constants. The crucial requirement for
integrability is the relative coefficient between the permutation
and trace operators which has to be $(1-\frac{d}{2})$ for an $so(d)$
spin
chain. This clearly indicates the lack of integrability of the energy operators ({\ref{bose1},{\ref{bose2}).\\
{\bf Integrable Sub-sectors:}\\
Even when an $so(d)$ spin chain is not integrable it is possible to
find integrable sub-sectors of the theory. As has been reviewed in
the appendix, if one constructed and $su(d')$ integrable spin chain
out of the relevant $\mathcal{R}$ matrix the integrable nearest
neighbor Hamiltonian would be \beq H = \alpha
\sum_lP_{l,l+1}\label{heis}.\eeq $so(d=2d')$ or $so(d=2d'+1)$ of
course contains $su(d')$. If we focussed only on spins transforming
under this $su(d')$, then at the level of the Hamiltonian, it would
correspond to dropping the trace operator, and we shall be left with
an integrable $su(d')$ sector of the theory. To find this embedding
of $su(d')$ at the level of the Hilbert space, one needs to  isolate
a subspace of the Hilbert space of the spin chain such
that\\
{\bf 1:} It is annihilated by the trace operator\\
{\bf 2:} It remains closed under the action of the spin chain
Hamiltonian.\\
When $d=2d'$, one can always find an integrable $su(d')$ sub-sector
for the theory as follows. One can form $d'$ complex combinations of
the $2d'$ spins $i_1\cdots i_{2d'}$ as \beq \mathcal{I}_1 = i_1 +
\sqrt{-1} i_{d'+1} \cdots \mathcal{I}_{d'} = i_{d'} +
\sqrt{-1}i_{2d'}.\eeq The  sub-space of the spin-chain Hilbert space
spanned by states such as\beq |f> = f_{l_1\cdots
l_J}|\mathcal{I}_{l_1}\cdots \mathcal{I}_{l_J}>\eeq involve only
traceless $so(d)$ tensors and are hence annihilated by the trace
operator. In this sub-space, the spin chain Hamiltonian involves
only the identity and the permutation operators and is hence
integrable. When $d=2d'+1$, the biggest closed integrable sector is
the same as the case when $d=2d'$. Of course, we cannot use this
simplification to diagonalize any part of the Hamiltonian
({\ref{bose2}), for which we shall resort to some approximate
methods described later in the paper. However, the presence of the
integrable $su(d')$ sector allows us to study the spectrum of
Bosonic membranes in flat backgrounds in dimensions greater than
four.

There is an important exception to the list of integrable spin
chains formed out of $so(d)$ invariant $\mathcal{R}$ matrices
discussed above and that corresponds to $d=2$. In four dimensions,
which corresponds to the case $d=2$, the above analysis of
integrability does not apply and  the corresponding one-loop spin
chain does turn out to be integrable! Since the Bethe ansatz for
this particular case is rather transparent and the generic nature of
the spectrum of the $d=2$ case is indeed the same as that of the
spin chains corresponding to integrable $su(d')$ subsectors of
membranes in higher dimensions we shall start by analyzing this
special case first. We shall then  generalize the result to the
integrable $su(d')$ sectors of various bosonic membranes at one
loop.
\subsection{The Special Case of $d=2$} To solve for the spectrum of  the special cased $d=2$,
the one loop energy operator({\ref{bose1}) we shall employ the
techniques of coordinate space Bethe ansatz.  Let us  first write
the trace and permutation operators in terms of the Pauli spin
matrices as follows. \beqs P_{l,l+1} &=&
 \frac{1}{2}\left( I(l)\otimes I(l+1) + \sum _i \sigma ^i(l)\otimes
 \sigma ^i(l+1)\right)\nonumber \\
 K_{l,l+1} &=& \frac{1}{2}\left( I(l)\otimes I(l+1) +
 \sigma ^x(l)\otimes \sigma ^x(l+1) +  \sigma ^z(l)\otimes \sigma
 ^z(l+1) -  \sigma ^y(l)\otimes \sigma ^y(l+1)\right).\eeqs
 The energy operator can then we expressed as
 \beq \Delta_{1,so(2)} = \frac{1}{8\mu^2}\left(\sum_{l=1}^J[\sigma^x(l)\sigma^x(l+1)
+\sigma^z(l)\sigma^z(l+1)
+3\sigma^y(l)\sigma^y(l+1)]\right)\label{spinh}\eeq This  is nothing
but the anisotropic Heiseberg chain, also known as the $XXZ$ spin
chain\footnote{The name the model refers to the fact that the
direction of the anisotrpy is usually chosen to be the $Z$
direction. We can bring the model to that form simply by re-labeling
the sigma matrices.}. The spin chain has an obvious symmetry which
allows one to carry out rotations in the $x-z$ plane by rotating
$\sigma^x$ and $\sigma ^z$ among each other. This is nothing but a
reflection of the original $so(2)$ symmetry of the interaction
$Tr[X_1,X_2]^2$ term of the matrix model Hamiltonian. We can also
express the hamiltonian in the form \beq \Delta_{1,so(2)} =
\gamma\left(\sum_{l=1}^J[S^1_2(l)S^2_1(l+1) +S^2_1(l)S^1_2(l+1)
+\delta
(S^1_1(l)S^2_2(l+1)+S^2_2(l)S^1_1(l+1))]\right)\label{spinw}\eeq In
the above equation, $\gamma = \frac{1}{4\mu^2}$ and $\delta = -3$.
We have introduced the Weyl operators, $S^1_j(l)$, that check if the
spin at the $l$ th site is equal to $j$ and if it is, then
$S^i_j(l)$ replaces the value of the spin by $i$. If the value of
the spin at the  $l$ th site does not match with $j$ then this
operator annihilates the  state. These operators satisfy the
associative Weyl agebra\beq S^i_j(l)S^k_m(l) = \delta
^k_jS^i_m(l)\eeq at the same site and the Weyl operators at
different sites commute with each other. With  this form
(\ref{spinw}) of the Hamiltonian, it is clear that the Ferromagnetic
state, with all the spins up(1) is a ground state of the
Hamiltonian.\beq |0> = |1,1,1\cdots 1>\eeq zero energy. The complete
set of excited states of the Hamiltonian can be built by considering
linear combinations of states with a fixed number of down/impurity
spins. The one magnon state or the state with a single impurity can
be taken to be of the form\beq |\Psi_1> = \sum_{l=1}^Jf(l)|l>,\eeq
where $|l>$ is the state with an upturned spin at the $l$th site. If
the state is taken to be a plane wave\beq f(l) = e^{ipl}\eeq then it
is easy to see that it will be a eigenstate of the Hamiltonian with
eigenvalue\beq E(p) =
-4\gamma\left(1+2\sin^2(\frac{p}{2})\right).\eeq The allowed values
of the momenta are determined from the condition that the state be
cyclically symmetric i.e \beq e^{ipJ} = 1 \Rightarrow p =
\frac{2n\pi}{J}.\label{dis}\eeq The one magnon state gives us the
dispersion relation (\ref{dis}) and we shall have to consider the
two magnon state to get the two magnon scattering matrix. The state
with two 'down' spins can be written as\beq |\Psi_2> =
\sum_{(x_1<x_2)=1}^J f(x_1,x_2)|x_1,x_2>\eeq where it is understood
that $|x_1,x_2>$ is the state with upturned spins at $x_1$ and
$x_2$. Once again we can make a Bethe ansatz by taking a linear
combination of plane waves that interact with each other by
exchanging momenta\beq f(x_1,x_2) = e^{i(p_1x_1+p_2x_2)} +
\mathcal{S}(p_1,p_2)e^{i(p_2x_1+p_1x_2)}.\eeq Such a state will be
an eigenstate of the Hamiltonian if the two magnon scattering matrix
$\mathcal{S}(p_1,p_2)$ is given by\beq \mathcal{S}(p_1,p_2) =
-\frac{ 1+ e^{i(p_1+p_2)} + 2\delta e^{ip_2}}{1+ e^{i(p_1+p_2)} +
2\delta e^{ip_1}}.\label{smat}\eeq The eigenvalue is given by\beq
E(p_1,p_2) = -4\gamma\sum_{i=1,2}\left(1 +
\sin^2(\frac{p_i}{2})\right),\eeq which shows that the energies of
the two magnons simply add up. The momenta are  determined by the
condition that the total phase picked up by one of the magnons by
travelling around the spin chain is nothing but the two body
scattering matrix. i.e.\beq e^{ip_1J} =\mathcal{S}(p_1,p_2),
e^{ip_2J} =\mathcal{S}(p_2,p_1).\eeq These are the Bethe equations
for the two unknowns $p_1,p_2$ and they provide us with the complete
solution to the two magnon problem. The $XXZ$ quantum spin chain is
an integrable system, and one of the manifestations of integrability
is the factorized nature of the multi-magnon scattering matrix into
two particle $S$ matrices. In other words, the above equations also
provide us with the solution to the the $m<J$ magnon problem. The
eigenstate with $m$ down spins has energy given by \beq E = -4\gamma
\sum_{i=1}^m \left(1 + \sin^2(\frac{p_i}{2})\right),\label{so2d}\eeq
with the momenta determined by\beq e^{ip_kJ} = \prod_{(j\neq
k)=1}^m\mathcal{S}(p_k,p_j).\label{bae}\eeq These are $m$ equations
for the $m$ unknown momenta that provide us with the complete
spectrum of the spin chain.  To get a feel for the qualitative
nature of the spectrum as a function of $J$ we can invoke a $BMN$
limit and work in a the dilute gas approximation, where the number
of magnons is very small compared to the length of the chain i.e
$m<<J$. We can write(\ref{bae}) by taking the logarithm of both
sides of the equation as \beq p_kJ = 2n\pi + \Theta(p_1,p_2).\eeq
$\Theta $ is the two body scattering phase shift. In the dilute gas
approximation, the magnons do not scatter at all, and have $\Theta
\sim 0$. The dispersion relation for the magnons can then be written
as \beq E = -4\gamma \left(\sum_{i=1}^m \left(1 +
\left(\frac{n_i\pi}{J}\right)^2\right)\right).\eeq This is the
leading order behavior of the spectrum of the  theory. There will in
general be corrections to this formula involving higher powers of
$\frac{1}{J}$. However if we take the limit, where,
$\frac{\lambda}{J^2}$ is small i.e the BMN limit\cite{bmn},  then
the above formula, which resulted from a one loop large $N$
perturbative computation should be expected to capture the
qualitative nature of the spectrum of the  theory to leading order
in $J$.
\subsection{Bethe Ansatz for the Integrable $SU(d')$ Sub-Sectors}
The large $N$ one loop energy operator in the $su(d')$ sector of the
spin chains (\ref{bose1}) can be written as (\ref{heis}) with
$\alpha = -\frac{1}{2\mu^2}$, as the restriction to the $su(d')$
sector simply amounts to dropping the trace operator from the spin
chain Hamiltonian.  The spins for this chain are in the fundamental
representation of $su(d')$. This spin chain is known in the
literature as the generalized Heisenberg model. It is known to be
integrable and the corresponding Bethe equations take on the form
\beq \left( \frac{u_{m,i} + i\vec{\alpha } _{m}. \vec{w}}{u_{m,i} -
i\vec{\alpha }_{m}.\vec{w}}\right)^J= \prod_{j\neq
i}^{n_m}\frac{u_{m,i} -u_{m,j}+ i\vec{\alpha } _{m}.
\vec{\alpha}_m}{u_{m,i} -u_{m,j}- i\vec{\alpha } _{m}.
\vec{\alpha}_m}\prod_{m'\neq m}\prod_{j\neq i}^{n_{m'}}\frac{u_{m,i}
-u_{m',j}+ i\vec{\alpha } _{m}. \vec{\alpha}_{m'}}{u_{m,i}
-u_{m',j}- i\vec{\alpha } _{m}. \vec{\alpha}_{m'}}\eeq $\alpha _m$
are the simple roots of the Lie algebra and $w$ is the highest
weight of the representation. The details of the derivation of these
Bethe equations have been reproduced in the appendix. To use the
Bethe equations and derive the results that put this generic case on
the same footing as the previous results on the special case of the
$so(2)$ case let us analyze the case of the above equation
corresponding to all the impurities of the $1$ type. In general, one
can have as many types of impurities as the rank of the Lie algebra.
However, it is enough to study the special case $m=1$ to get a feel
for the nature of the spectrum. In this case, calling$u_{1,i}$
$\mu_i$, the Bethe equations become\beq \left(\frac{\mu_j +
i/2}{\mu_j - i/2}\right)^J = \prod_{k\neq j}\frac{\mu_j -\mu_k +
i/2}{\mu_j -\mu_k - i/2}\eeq By writing the equation in terms of the
momenta $p_i$ \beq e^{-ip} = \frac{\mu - i/2}{\mu + i/2},\eeq rather
than the rapidities, we can write the Bethe equations as \beq
e^{ip_kJ} = \prod_{(j\neq k)=1}^m\mathcal{S}(p_k,p_j),\eeq where
\beq \mathcal{S}(p_1,p_2) = -\frac{ 1+ e^{i(p_1+p_2)} + 2
e^{ip_2}}{1+ e^{i(p_1+p_2)} + 2e^{ip_1}}.\eeq This is completely
analogous to the equations for the $so(2)$ spin chain, except that
$\delta =1$ in the present case. That is to be expected, because,
the case of having impurities of the $1$ type reduces the problem to
the usual spin half Heisenbrg chain, which is the same as the $XXZ$
chain with the value of the anisotrpy $\delta =1$. As discussed in
the appendix, the dispersion relation takes on the form\beq E=
\sum_{i=1}^m \epsilon(p_i), \epsilon(p) = -\frac{1}{\mu^2}\cos(p).
\eeq As before, one can also take the limit of large $J$, in which
case the magnons/impurities behave as free particles on a circle. In
this limit the momenta are given by \beq p_i =\frac{2n_i\pi}{J},\eeq
where $n_i$ are integers. They can take on any value, but must be
constrained to obey the level matching or zero momentum constraint
implied by the cyclicity of the trace i.e $\sum_i n_i = 0$. The
dispersion relation, simplifies to \beq \epsilon(n) = -\left(1 +
\frac{2n^2\pi^2}{J^2}\right).\eeq If one allowed for other types of
impurities, then the form of the Bethe equations would be more
complicated, however, the qualitative facts about the nature of the
spectrum given by the simplest case above do not change. The effect
of the other types of impurities, in a chain of finite length would
be to change the allowed values of the momenta. However, the
dispersion relation would remain the same. Moreover, only $u_{1}$
enters the dispersion relations, thus the only effect that the other
types of impurities would have are indirect.  Although the Bethe
equations written out in their full form are rather complicated
algebraic equations, the lessons learnt from the simplest case of
$1$ type of impurities carry over to the more general scenario as
well.
\subsection{Enhancement of Integrability in the large $J$ Limit}
From the previous discussion of the construction of integrable
$so(n)$ invariant spin chains with nearest neighbor interactions
from the the standard $\mathcal{R}$ matrix it follows that the form
of the Hamiltonian is highly constrained from the requirements of
integrability. Only Hamiltonians with very special relative
coefficients between the permutation and the trace operator fulfill
this requirement. Although, we were able to find rather large
integrable subsectors for the $so(d)$ spin chains, it is worthwhile
to ask the question whether or not one can find integrable behavior
in some limit of the spin chains without having to resort to the
truncation to sub-sectors. If it is indeed possible, then we should
be able to gauge the behavior of the complete spectrum of the theory
albeit in some appoximation scheme. The simplification of the nature
of the spectrum in the large $J$ limits of the various exactly
solvable sectors described in the previous sections suggests that
the large $J$ limit might be such an approximation. This is indeed
the case. In this final section on the Bosonic matrix models of the
paper, we shall formally demonstrate that the large $J$ limits of
the various $so(d)$ spin chains studied so far can be mapped to
integrable non-linear sigma models irrespective of whether the
original spin chain is integrable or not.

Since we are interested in spin chains with nearest neighbor
interactions, and nearest neighbor interactions are nothing but the
lattice Laplacian, the non-linear sigma models we expect to
approximate the low lying spectrum of the spin chains will be free
sigma models with an appropriate homogeneous space as its target
space. However, most classical sigma models of this kind are known
to be integrable. Thus, it is reasonable to expect that the low
lying excitations of an $so(d)$ invariant quantum spin chain with
nearest neighbor interactions will be described by an integrable
classical sigma model even if the original spin chain is {\it{not}}
integrable. We shall now proceed to make this idea more precise. We
begin by writing the $\mathcal{R}$ matrix in a slightly different
but equivalent form \beq \mathcal{R}(u)_{i,j} = I_{i,j} +
\frac{\hbar }{u}P_{i,j} + \frac{\hbar }{\hbar g'-u}K_{i,j}, g' =
\left( 1 - \frac{d}{2}\right)\label{rson}. \eeq $P_{i,j}, K_{i,j}$
are the operators that permute and trace over the spins belonging to
the $i$ and $j$th vector spaces respectively. $\hbar $ is the
natural `quantum' deformation parameter in the problem. An
alternative way of writing the $R$ matrix is as a $d \times d$
operator valued matrix. For example, $\mathcal{R}_{i,j}$ can be
expressed as a matrix, each of whole elements is an operator on the
$i$ th vector space as, \beq \mathcal{R}(u)^{i}_{\mu \nu } = \delta
_{\mu \nu} I(i) + \frac{1}{u}S_{\nu \mu}(i) + \frac{1}{g'-u}S_{\mu
\nu }(i), \mu, \nu = 1 \cdots d. \eeq In writing the $R$ matrix in
the second form, we have introduced the Weyl operators $S_{\mu, \nu
}(i)$, which act as $|\mu ><\nu|$ on the $i$ th vecotr space. The
Weyl operators satisfy the standard commutation relations, \beq
[S_{\mu \nu}(i), S_{\alpha \beta }(j)] = \hbar
\delta_{i,j}\left(\delta _{\nu \alpha }S_{\mu \beta} - \delta _{\mu
\beta}S_{\alpha \nu }\right). \eeq
 The relations satisfied by the transfer
matrix, the so called RTT relations, are written most succinctly as,
\beq R_{1,2}(u - v) T_1(u)T_2(v) = T_2(v)T_1(u) R_{1,2}(u - v). \eeq
$1,2$ refer to two auxilliary spaces. It is more insightful to write
this equation out in terms of the matrix elements of $T$. \beqs
\frac{1}{\hbar}[T_{ab}(u),T_{cd}(v)] + \frac{1}{u-v}\left(T_{ad}(u) T_{cb}(v) - T_{ad}(v)T_{cb}(u)\right)+\nonumber \\
\frac{1}{\hbar g' - (u-v)}\left(\delta_{a,b}T_{lc}(u)T_{ld}(v) -
\delta _{c,d} T_{al}(v)T_{bl}(u)\right) = 0 \eeqs We shall now write
the corresponding $RTT$ relation in the large $J$ limit. To do that
we shall define \beq\hbar = \frac{\hbar '}{J}, u',v' = \frac{u}{J},
\frac{v}{J}, \mathcal{S} = \frac{S}{J}\eeq In terms of the rescaled
variables, the $\mathcal{R}$ matrix becomes \beq
\mathcal{R}(u)^{i}_{\mu \nu } = \delta _{\mu \nu} I(i) + \frac{\hbar
'}{Ju'}\mathcal{S}_{\nu \mu}(i) +
\frac{\frac{\hbar'}{J}}{\frac{\hbar'g'}{J^2}-u'}\mathcal{S}_{\mu \nu
}(i)\approx \delta _{\mu \nu} I(i) + \frac{\hbar
'}{Ju'}\mathcal{S}_{\nu \mu}(i) + \frac{\hbar
'}{Ju'}\mathcal{S}_{\mu \nu }(i) + \mathcal{O}(\frac{1}{J^2}).\eeq
Needless to say that the $\mathcal{S}$ operators satisfy\beq
[\mathcal{S}_{\mu \nu}(i), \mathcal{S}_{\alpha \beta }(j)] =
\frac{\hbar '}{J} \delta_{i,j}\left(\delta _{\nu \alpha
}\mathcal{S}_{\mu \beta} - \delta _{\mu \beta}\mathcal{S}_{\alpha
\nu }\right). \eeq From the two preceding equations it is quite
clear that the term in the $R$ matrix that involved $g'$, and which
was ultimately responsible for the $g'$ dependence of the
Hamiltonian, is lower order in $\frac{1}{J}$ and that $\frac{\hbar
'}{J}$ plays the role of the deformation parameter in the large $J$
limit. The $RTT$ relations can be written out as\beqs
\frac{1}{\frac{\hbar'}{J}}[T_{ab}(u'),T_{cd}(v')] + \frac{1}{u'-v'}\left(T_{ad}(u') T_{cb}(v') - T_{ad}(v')T_{cb}(u')\right)+\nonumber \\
\frac{1}{(u'-v')}\left(\delta_{a,b}T_{lc}(u')T_{ld}(v') - \delta
_{c,d} T_{al}(v')T_{bl}(u')\right) = 0 \eeqs We also observe
that\beq[T_{ab}(u'),T_{cd}(v')] =
\frac{\hbar'}{J}\{t_{ab}(u'),t_{cd}(v')\} +
\mathcal{O}(\frac{1}{J^2})\eeq where $t$, as a matrix is the same as
$T$, except that its entries are to be thought of as ordinary
functions and not operators. Thus to leading order in $\frac{1}{J}$,
the $RTT$ relations may be replaced by \beqs
\{t_{ab}(u'),t_{cd}(v')\}+ \frac{1}{u'-v'}\left(t_{ad}(u') t_{cb}(v') - t_{ad}(v')t_{cb}(u')\right)+\nonumber \\
-\frac{1}{(u'-v')}\left(\delta_{a,b}t_{lc}(u')t_{ld}(v') - \delta
_{c,d} t_{al}(v')t_{bl}(u')\right) = 0, \eeqs This simply implies
that the matrix elements of $t$ satisfy the following Poisson
brackets: \beq \{t(u')\stackrel{\otimes}{,} t(v')\} = [r(u'-v'),
t(u') \otimes t(v')], \label{crtt}\eeq where, \beq r(x-y) =
\frac{1}{x-y}(P - K). \label{liep}\eeq To clarify the notation it is
worth mentioning that equation (\ref{crtt}) is to be thought of as
expressing the equality of the actions of both sides of the equation
on two copies of a $d$ dimensional vector space\cite{fadbook}. The
main implication of (\ref{crtt}) is that $\mbox{Tr}t$ is the
generator of an infinite family of conserved charges i.e \beq
\{\mbox{Tr}t(u), \mbox{Tr}t(v)\} =0.\eeq Our analysis of large $J$
integrability of the continuum sigma models obtained from
(\ref{bose1},\ref{bose2}) would be complete if we can show that the
relevant sigma models are contained in the family of commuting
charges that follow from the classical transfer matrix $t$. This is
indeed the case. From a technical point of view, taking the
continuum limit of the spin chain amounts to taking its coherent
state expectation value in a long coherent state of length $J$ and
deriving an effective Hamiltonian in terms of the group parameters
characterizing the coherent state. For $so(d)$ valued spin chains
with nearest neighbor interactions this has been worked out in
detail in\cite{ar1}, and we shall refer to that paper for the
detailed descriptions of many of the results that will be used here.
The coherent state construction of\cite{ar1} when applied to
(\ref{bose1},\ref{bose2}) generate the following sigma model
Hamiltonins \beq H = \alpha\int
dx\mbox{Tr}\left(\partial_xM\partial_xM\right).\label{sham}\eeq
$\alpha $ is the effective coupling of the sigma model. It is
$-\frac{1}{8\mu^2J^2}$ for (\ref{bose1}) and  $\frac{3\theta^2
-1}{8\mu^2J^2}$ for (\ref{bose2}). $x$ is nothing but the contunuum
coordinate along the length of the spin chain. $M(x,t)$ is a
$d\times d$ antisymmetric matrix and it satisfies the Poisson
brackets \beq \{M(x)_{ij}, M(x')_{kl}\} =
\delta(x-x')(\delta_{jk}M_{il}(x)+\delta_{il}M_{jk}(x)-\delta_{ik}M_{jl}(x)-\delta_{jl}M_{ik}(x))\eeq
The matrix $M$ is built out of group parameters that parameterize
the $so(d)$ coherent state and it can be written in terms of complex
$d$ vectors $Z_i$ that satisfy \beq Z_iZ_i = Z^*_iZ^*_i = 0,
Z^*_iZ_i =1\eeq as \beq M_{ij}= Z_iZ^*_j-Z_jZ^*_i. \eeq This
parametrization along with the various various algebraic identities
for $M$ that it leads to have also been discussed in\cite{ar1}. The
equations of motions for $M$ can be derived using the Poisson
structure given above\beq
\partial_tM = 4\alpha \partial_x[M,\partial_xM]\eeq The equations of motion are equivalent
to the following flatness condition. \beq
[\partial_t+\mathcal{A}_t,\partial_x + \mathcal{A}_x] = 0,\eeq where
\beqs \mathcal{A}_x = \frac{1}{u}M\nonumber \\
\mathcal{A}_t = \frac{4\alpha}{u}[M,\partial_xM] -
\frac{4\alpha}{u^2}M\eeqs are the two components of the Lax
connection.  The Poisson brackets between the spatial components of
the Lax connection can be written down as\beq \{\mathcal{A}_x(x,u),
\mathcal{A}_x(x',u')\} = \delta(x-x')[r(u-u'),
\mathcal{A}_x(x,u)\otimes I + I\otimes \mathcal{A}_x(x',u')].\eeq
The above equation expresses the Poisson bracket between the
dynamical variables as a Lie bracket in $so(d)$. This is the
familiar Lie-Poisson structure characteristic of classical
integrable two dimensional field theories\cite{fadbook}. The
immediate consequence of the Lie-Poisson structure is that \beq t(u)
= P e^{\int_0^J \mathcal{A}_x(x)dx}\eeq satisfies the Poisson
brackets (\ref{crtt}). This  of course implies the fact that
$\mbox{Tr}t(u)$ is a generating function of the integrals of motion
for the Hamiltonian (\ref{sham}).  This establishes the
integrability of the sigma model describing the large $J$ limit of
the spin chain. We thus see that {\it all} the nearest neighbor spin
chains that arise as the one loop energy operators for matrix models
with dynamically generated masses correspond to integrable sigma
models in the large $J$ limit.

A cautionary remark is in order while discussing the continuum
limits of the spin chains(\ref{bose1}, \ref{bose2}). The $so(d)$
invariant spin chain (\ref{bose1}) is Ferromagnetic and the
continuum limit described above does approximate the spectrum around
the Ferromagnetic ground state. However, (\ref{bose2}) is
anti-ferromagnetic. Thus it is to be understood that the sigma model
following from the above construction, when applied to the $so(3)$
case  approximates the spectrum of the spin chain around its highest
energy state. The  sigma model following from the continuum limit of
the $so(3)$ chain around the anti-Ferromagetic vacuum is a level two
$su(2)$ Wess-Zumino model, and we shall refer to\cite{wzw1,wzw2} for
the details of that continuum limit.

\section{Mass Deformed Super-Membranes and Integrability:} In the
final section we shall analyze models of mass deformed
supersymmetric quantum mechanics obtained from the dimensional
reduction of minimally supersymmetric Yang-Mills theories in various
dimensions. The resultant matrix models can be taken to be matrix
regularized light cone descriptions of super-membranes propagating
in plane wave type backgrounds in one higher dimension than the one
in which the original gauge theory was formulated. The gauge
theories in question are minimally supersymmetric Yang-Mills
theories in spacetime dimensions six four and three. If one carried
out a naive dimensional reduction of these Yang-Mills theories to
one dimension, the resultant matrix quantum mechanics system would
not have any quadratic mass terms. The matrix quantum mechanics
models can nevertheless be interpreted as regularized versions of
non-critical M(atrix) theory. This is very much along the lines of
the original BFSS proposal for M(atrix) theory in eleven dimensions.
The gauge theory in question there is of course $\mathcal{N}=1$ SYM
in $D=10$. In the presence of enough supersymmetries, one cannot be
confident of the success of the mechanism for dynamical mass
generation that we applied to the case of Bosonic membranes earlier
in the paper\footnote{In a recent paper\cite{mem-gap} it has been
argued that if one allowed for central extensions of the underlying
SUSY algebra for the matrix models then mass gaps might be generated
even in the presence of supersymmetry. It would of course be
interesting to see if the methods of resummed perturbation theory
can capture the mass gaps of systems posessing centrally extended
supersymmetry}. There might be 'true' flat directions in the quantum
mechanical system due to cancelations between Bosonic and Fermionic
self-energy diagrams.\footnote{Even for supersymmetric matrix
models, one can apply re-summed perturbation theory to estimate mass
gaps at finite temperatures. Indeed, such an analysis for the eleven
dimensional M(atrix) theory Hamiltonian has been carried out by
Kabat and collaborators in\cite{kl}. It might thus be possible to
set up a systematic perturbation theory using a combination of
spin-chin techniques and the mechanism of dynamical mass generation
in the case of supersymmetric matrix models at finite temperatures
as well.} However, there does exist a second mechanism for adding
mass terms in supersymmetric matrix models. One can add explicit
mass terms to the matrix model Hamiltonian and ask for for a
preservation of all the supersymmetries of the un-deformed model.
This greatly constrains the permissible types of mass deformations,
and all possible mass deformations of the matrix models obtained
from dimensional reductions of minimally supersymmetric Yang-Mills
theories in various dimensions were carried out in\cite{kp}.
Sometimes, it is possible to interpret the massive matrix quantum
mechanics as being related to the dimensional reduction of gauge
theories, not on $R^m$ but on on $R\times M^{m-1}$, where $M$ is a
$m-1$ dimensional compact manifold. For example the BMN matrix model
can be regarded as the dimensional reduction of $\mathcal{N}=4$ SYM
on $R\times S^3$\cite{bmn=4}. Obviously, the question as to whether
or not there is always such an interpretation of massive
supersymmetric matrix quantum mechanical systems possibly requires
further study. Apart from the connection to Yang-Mills theories,
such mass deformed matrix models can be regarded as light-cone
Hamiltonians for super-membranes in plane wave like backgrounds,
once again, in one dimension higher than the dimensionality of the
original Yang-Mills theory\cite{kp}. Thus, they are quite naturally
models for non-critical M(atrix)theory Hamiltonians. Thought of in
another way, this is nothing but the extension of the relation of
the BMN matrix model to a critical supermembrane
theory\cite{keshav1} in a plane wave background to non-critical
dimensions. Taking a clue from the recently discovered integrable
behavior in the BMN model\cite{bmn=4} it is natural to ask, whether
or not the matrix models for non-critical membrane correspond to
integrable spin chains in the large $N$ limit. That is the problem
that we shall concern ourselves with in this part of the paper. As
in the BMN model\cite{bmn=4}, a good starting point for the anlysis
of integrability is provided by the Bosonic sectors of these models
at one loop. Without integrabilty in the Bosonic sector at the one
loop level, there is of course no hope of expecting integrability at
higher loops. In what follows, we shall systematically analyze the
Bosonic sectors of all  models proposed by Kim and Park\cite{kp} at
the one loop level at large $N$.

The various mass deformed models obtained by Kim and Park\cite{kp}
can be classified into two broad
catagories.\\
{\bf I:} Models that admit non-trivial supersymmetric configurations
saturating the unitary bounds of the corresponding super algebras.\\
{\bf II:} Models that do not posses any non-trivial supersymmetric
configurations\\
In what is to follow, we first catalogue all the bosonic subsectors
of the models presented in\cite{kp}. We then go on to show that all
the models that admit supersymmetric vacua also correspond to
integrable spin chains at the one loop level. On the other hand,
models that do not have non-trivial supersymmetric vacua, do not
correspond to spin chains that are integrable, but they do possess
integrable sub-sectors.
\subsection{$\mathcal{N}$ =8 Super Matrix Quantum Mechanics}
As mentioned before, the massive matrix model obtained from
$\mathcal{N}=1$ SYM in $D=10$ is nothing but the BMN matrix model
and it has been studied in substantial detail in the recent
past\cite{bmn=4}. The theory involves sixteen supercharges, and it
shares many common features with $\mathcal{N}=4$SYM on $R\times
S^3$. Among the list of common features is the integrability of a
rather large sub-sector of the matrix model, the so called $su(2|3)$
sub-sector to the third order in large $N$ perturbation
theory\cite{su2/3}. At the one loop order, the complete theory,
without any truncation to specific sub-sectors also exhibits
integrability\footnote{An exhaustive account of the one-loop
integrability of the BMN matrix model along with its relation to
$\mathcal{N}=4$SYM can be found in\cite{klose1}}. The next lower
dimension that admits a consistent formulation of $\mathcal{N}=1$
SYM is six. The mass deformed matrix quantum models obtained from
this Yang-Mills theory are of two types. The matrix model that has
non-trivial susy vacuum solutions has as its Bosonic part
 \beq H = \mbox{Tr}\left( \frac{1}{2}\Pi_i\Pi_i +
\frac{1}{2}(\frac{\mu}{3})^2 X_aX_a +
\frac{1}{2}(\frac{\mu}{6})^2X_AX_A -\frac{1}{4}[X_i,X_j]^2 +
i\mu[X_3,X_4]X_5 \right)\eeq $a= 3,4,5$ and $A=1,2$, while $i,j$ in
the quartic interaction term run from $1\cdots 5$ in the above
equation. The mass deformation breaks the $SO(5)$ symmetry of the
original gauge theory down to $SO(3)\times SO(2)$. This models
admits a supersymmetric configuration corresponding to circular
motion in the $(1,2)$ plane and a fuzzy sphere in the $(4,5,6)$
directions. \beq X_1 = R\mbox{cos}(\frac{1}{6}t\mu)\mathcal{I}, X_2
= R\mbox{sin}(\frac{1}{6}t\mu)\mathcal{I}, X_a = \frac{1}{3}\mu
\mathcal{J}_a\eeq $\mathcal{J}$ are the standard $so(3)$ generators
satisfying\beq [\mathcal{J}_a,\mathcal{J}_b] =
i\epsilon_{abc}\mathcal{J}_c.\eeq Clearly, states that involve only
Bosonic excitations are eigenstates of the free part of the
Hamiltonian, which is a sum of decoupled harmonic oscillators. It
thus makes sense to compute the corrections to the tree level
energies of the Bosonic excitations of this model. To do such a
computation at the one loop level, one will have to take into
account the contributions from the Fermionic part of the
Hamiltonian, since the Fermions do run in loops. However, as far as
the one-loop energy operators is concerned, the analysis of the BMN
matrix models shows that contribution of the Fermionic terms to the
one loop effective Hamiltonian, when written down in the spin chain
language is proportional to the identity
operator\cite{bmn=4,klose1}. But since we are interested in
measuring the energies with respect to  a reference state, that
contribution of the Fermions to the energy operator as we have
defined it in the paper vanishes. This general observation is true
of the present  and all the other models that we study in this
section. Since the symmetry algebra of the matrix model is
$so(3)\times
 so(2)$, we can read off the contributions of these sectors to
$\Delta ^1$, using the basic relations(\ref{ensod},\ref{enso3}).
 \beq \Delta ^1_{so(2)} = \frac{18}{\mu ^2}\sum_l
\left( \frac{1}{2}K_{l,l+1} - P_{l,l+1}\right).\eeq \beq\Delta
^1_{so(3)} = \frac{9}{2\mu ^2}\sum_l \left(2P_{l,l+1}
-4K_{l,l+1}\right).\eeq The dimensional reduction of six dimensional
super Yang-Mills theory also admits a second type of mass
deformation, where the bosonic $so(5)$ symmetry is broken to
$so(4)\times u(1)$. The Bosonic Hamiltonian for this mass
deformation is\beq H = \mbox{Tr}\left(\sum_{i=1}^5\Pi_i\Pi_i +
\frac{1}{2}(\frac{\mu }{3})^2X_1X_1 +\frac{1}{2}(\frac{\mu
}{6})^2\sum_{A=1}^4X_AX_A -\frac{1}{4}\sum_{a<b
=1}^5[X_a,X_b]^2\right)\eeq In this mass deformation, there are no
BPS solutions analogous to the fuzzy sphere solution of the
$so(3)\times so(2)$ case. When one looks at $\Delta ^1$ for this
model, the only non-trivial contribution to it comes from the
$so(4)$ part. As before, using (\ref{ensod}) we can write down \beq
\Delta ^1_{so(4)} = \frac{18}{\mu ^2}\sum_l \left(
\frac{1}{2}K_{l,l+1} - P_{l,l+1}\right).\label{delso4}\eeq The spins
are now in the defining representation of $so(4)$.
\subsection{$\mathcal{N}$ =4 Super Matrix Quantum  Mechanics}
Next in our catalog, are the matrix models obtained from a mass
deformation of the dimensional reduction of minimal super Yang-Mills
in $D=4$. As in the previous case, one can do two consistent mass
deformations\cite{kp}.  The Bosonic part of the Hamiltonian that
admits a maximally supersymmetric fuzzy sphere configuration is\beq
H = \mbox{Tr}\left( \frac{1}{2}\Pi_i\Pi_i +
\frac{1}{2}(\frac{\mu}{3})^2 X_aX_a -\frac{1}{4}[X_i,X_j]^2 +
i\mu[X_1,X_2]X_3  \right)\eeq  This matrix model corresponds to the
$so(3)$ symmetric part of the previous model and the corresponding
fuzzy sphere vacua are also the same.\beq X_a = \frac{1}{3}\mu
\mathcal{J}_a, [\mathcal{J}_a,\mathcal{J}_b] =
i\epsilon_{abc}\mathcal{J}_c.\eeq The one loop energy operator for
the scalars of this model is also precisely the same as the one
obtained above. \beq \Delta ^1_{so(3)} = \frac{9}{2\mu ^2}\sum_l
\left(2P_{l,l+1} -4K_{l,l+1}\right).\eeq A second kind of mass
deformation of the dimensional reduction of $D=4,\mathcal{N}=1$ SYM
results in a Bosonic Hamiltonian that has the $so(3)$ symmetry
broken down to $so(2)\times u(1)$. The Hamiltonian for the Bosonic
part is \beq H = \mbox{Tr}\left( \frac{1}{2}\Pi_i\Pi_i +
\frac{1}{72}\mu ^2 (X_1X_1 + X_2X_2 + 4X_3X_3)
-\frac{1}{4}[X_i,X_j]^2 \right)\eeq This model too admits a
supersymmetric configuration preserving all of the four
supercharges\cite{kp}. The supersymmetric directions correspond to
\beq X_1 = R\mbox{cos}(\frac{1}{6}t\mu)\mathcal{I}, X_2 =
R\mbox{sin}(\frac{1}{6}t\mu)\mathcal{I}, X_3=0\eeq \beq \Delta
^1_{so(2)} = \frac{18}{\mu ^2}\sum_l \left( \frac{1}{2}K_{l,l+1} -
P_{l,l+1}\right).\eeq
\subsection{$\mathcal{N}$ = 2 Super Matrix Quantum Mechanics}We now
move on to the final case, which is that of dimensional reduction of
three dimensional minimal super Yang-Mills. Unlike the previous
cases,  there is a unique mass deformed matrix model that one can
obtain from minimal super Yang-Mills in three spacetime dimensions.
The Bosonic part of the Hamiltonian of this model is \beq H =
\mbox{Tr}\left( \frac{1}{2}\Pi_i\Pi_i + \frac{1}{72}\mu ^2 (X_1X_1 +
X_2X_2) -\frac{1}{4}[X_i,X_j]^2 \right)\eeq  There is a maximally
supersymmetric BPS configuration ins this case as well. It is  given
by the static solution\beq X_1 =
R\mbox{cos}(\frac{1}{6}t\mu)\mathcal{I}, X_2 =
R\mbox{sin}(\frac{1}{6}t\mu)\mathcal{I}.\eeq Finally, we write down
the one loop energy operator for the model, which  takes on the
form:\beq
 \Delta ^1_{so(2)} =
\frac{18}{\mu ^2}\sum_l \left( \frac{1}{2}K_{l,l+1} -
P_{l,l+1}\right).\eeq {\bf Integrability:}\\
Having listed all the bosonic one-loop energy operators for the mass
deformed models, we can just easily understand whether or not the
spin chains are integrable or not. The basic results are summarized
in the following table.
\begin{center}
\begin{tabular}{|c|c|c|}
  Supercharges & Symmetry & Integrable Sectors\\
  8 & $so(3)\times so(2) $& Both $so(3)$ and $so(2)$ sectors\\
  8 & $so(4)\times u(1)$ & $su(2)(\in so(4))$ sector\\
  4 & $so(3)$& $so(3)$ sector\\
  4 & $so(2)\times u(1)$ & $so(2)$ sector\\
  2 & $so(2)$& $so(2)$ sector\\
\end{tabular}
\end{center}
From the summary given above, we see that all the cases that do
support non-trivial fuzzy sphere type of supersymmetric
configurations give rise to integrable spin chains. The
corresponding spin chains have $so(3)$, $so(2)$ or their product
(corresponding to $\mathcal{N} =2,4$ and 8 respectively) as their
symmetry groups.   Integrability of the $so(2)$ sectors of the
$\mathcal{N}$$=8,4$ and$2$ cases follows directly from the
realization of $so(2)$ invariant nearest neighbor spin chains as
$xxz$ models discussed previously. We can simply use these results
to derive the Bethe ansatz for the three $so(2)$ cases of interest
in the supersymmetric context. This simply amounts to replacing $\mu
$ by $\frac{\mu}{6}$ in the formulae(\ref{so2d},\ref{bae}) for the
$so(2)$ Bethe ansatz given
previously.\\
As far as the two $so(3)$ invariant sectors of the $\mathcal{N}=8$
and $\mathcal{N}=4$ cases are concerned, they are integrable as
well!  From form of the most general integrable $so(d)$ spin chain
with nearest neighbor interactions given in(\ref{intsod}), we see
that the $so(3)$ chains given above have the correct form
commensurate with integrability. To get the spectrum and the Bethe
equations, one can use the known Bethe equations for the defining
representation of $so(3)$\cite{genh1,genh2,genh3,genh4}. Since
$so(3)$ has only one Cartan generator, the Bethe equations only
allow for one type of impurity and they are very similar to the
$so(2)$ equations. When written out in terms of the rapidities they
read as \beq \left(\frac{u_k +i}{u_k -i}\right)^J = \prod_{j\neq
k}\left(\frac{u_k - u_j +i}{u_k -u_j -i}\right),\eeq while the
dispersion relation is\beq \epsilon(u) =
-\frac{36}{\mu^2}\frac{1}{u^2+1}\eeq for  the $so(3)$ chains related
to both the $\mathcal{N}=8$ and $\mathcal{N} =4$ cases discussed
above. This provides the complete one-loop solution for the two
$so(3)$ sectors of interest.

The one example of a mass deformed model that does not have any
non-trivial supersymmetric vacua corresponds to the case of the
$\mathcal{N} =8$ matrix quantum mechanics with symmetry group
$so(4)\times u(1)$. The $so(4)$ spin chain (\ref{delso4}) is clearly
{\it not} integrable. However, as in the case of the matrix models
for Bosonic memebrane theories in flat spacetimes discussed
previously in the paper, one can find integrable $su(2)$ subsectors
of the theory by looking at states formed by   excitations
corresponding to $Z = X_1+iX_3$ and $W = X_2+iX_4$. The resultant
spin chain is nothing but the spin one-half Heisenberg model, whose
Bethe equations are \beq \left(\frac{u_k +i/2}{u_k -i/2}\right)^J =
\prod_{j\neq k}\left(\frac{u_k - u_j +i}{u_k -u_j -i}\right).\eeq
The total energy is given by \beq E = -\sum_i
\frac{36}{\mu^2}\left(\frac{u_i^2 - 1/4}{u_i^2 + 1/4}\right). \eeq
This completes the discussion of the one-loop spectra of the Bosonic
sectors of the models proposed in\cite{kp}. Whether or not, this
understanding of integrability extends to the more general sectors
of theme models for supermembranes and whether or not integrability
is preserved at higher loops are of course open questions that might
be of interest.

Before concluding the segment on the analysis of the matrix models
given in\cite{kp}, it is worth remarking that we have left out one
particular example that was also presented in\cite{kp}. This last
example corresponds to $\mathcal{N}=1$ matrix quantum mechanics and
is related to the dimensional reduction of minimal SYM in $D=2$.
This matrix model, which has a single Bosonic and a single Fermionic
degree of freedom cannot quite be studied using the formalism
presented here, as one needs at least two Bosonic matrices to be
able to make it work. However, a different approach, might be in
order here. Supersymmetric matrix quantum mechanics systems,
involving a single Bosonic matrix degree of freedom can often be
mapped to supersymmetric analogs of the Calogero-Sutherland model,
see for example\cite{acal}. It might be possible to utilize such a
connection in this case leading possibly to some exact statements
about this model. This possibility probably merits  further
investigation.
\section{Concluding Remarks}
In this final section, we briefly return to the analysis of Bosonic
membranes and sketch out some directions for future investigations.
In sections three and four we presented a mechanism for estimating
the non-perturbative mass-gaps in the spectra of Bosonic membranes
and also developed an expansion of the matrix model Hamiltonians
around the quantum corrected effective potentials using the
techniques of quantum spin chains. It is of course implied that we
were expanding the Hamitonian around the trivial/oscillator vacuum
around the effective potential. The low-lying excitations around the
effective potential were shown to be well described by closed spin
chains. In the D-brane picture, the closed spin chains could be
thought of as the low-lying closed string like excitations of the
membrane. Indeed, the sigma model for the spin chains derived as the
continuum limits of the spin chains can be thought of as an
effective string sigma model in a background provided by the
membrane. The membrane point of view then provides us with an
open-string interpretation of the sigma model. One could take this
interpretation seriously and ask if there are analogues of bona-fide
open string degrees of freedom in the membrane picture as well. This
it indeed possible, if one looked at particular $O(N)$ excitations
of the matrix models. The closed spin chains arose in our analysis
because we focussed on gauge invariant operators built out of
traces. However, taking a cue from the analysis of the dilatation
operator of $\mathcal{N}=4$ SYM, we could just as well have looked
at states built out of determinants and sub-determinants\cite{vij1}.
Such operators can also be studied within the paradigm of quantum
spin chains, except, one would have open spin chains i.e the analogs
of the open string degrees of freedom to consider. Issues related to
integrability of these spin chains tend to be rather subtle (see for
example\cite{aopen}) and it might be of interest to investigate
these degrees of freedom for bosonic membranes as well.

The complete spectrum of a theory of membranes is of course much
richer than what has been discussed in this paper. The richness of
the spectrum has to do with the fact that the trivial/oscillator
vacuum of the bosonic matrix models is only one of many other vacua
that are also present in the theories that we considered. An example
of a non-trivial vacuum in a matrix quantum mechanical system was
already alluded to in the section on supersymmetric matrix models.
These are the so called fuzzy sphere vacua. The Bosonic membranes,
coupled to Chern-Simons fluxes that were considered earlier in the
paper also possess such vacua\footnote{For a detailed study of
various vuzzy sphere vacua in matrix models of the type considered
in the paper see\cite{gm}}. These vacua continue to be present
around the quantum/mass corrected effective potentials as well. It
is well known that the expansions of the quantum mechanical matrix
models around fuzzy sphere vacua leads to gauge theories in $2+1$
dimensions. For instance, $D=2+1$, $\mathcal{N}=4$ SYM and its
connection to the expansion of the BMN matrix model around a fuzzy
sphere vacuum has been discussed at length in\cite{m30,m31}
\footnote{The connection between fuzzy/non-commutative spaces and
gauge theories has a rather large literature devoted to it. For a
pedagogical introduction we refer to\cite{szabo}.}. Put differently,
quantum mechanical matrix models can be thought of as providing a
regularized description of gauge theories in three dimensions. The
regularization is accomplished by introducing non-commutativity in
the spacial directions. Gauge theories in three dimensions of course
have mass-gaps in their spectra. For example, the mass gap in the
spectrum of pure Yang-Mils in three dimensions has been computed
in\cite{vp1,vp2,vp3,vp4,vp5}. It is thus plausible that the methods
for estimating spectral gaps for matrix quantum mechanics that were
presented in this paper can well be utilized to compute mass gaps
and glueball masses for $D=2+1$ Yang-Mills theories. Such a
possibility was also pointed out in\cite{mem-gap}. The mass gap in
$D=2+1$ Yang-Mills can be estimated in two different ways.
In\cite{vp1}, the gap was computed by applying re-summed
perturbation theory to the gauge theory, while in
\cite{vp2,vp3,vp4,vp5} a strong coupling expansion was utilized
within the Hamiltonian framework. The answers obtained from both the
approaches are strikingly close to each other. To draw a parallel at
a matrix model level, the analysis of\cite{mem-gap} is very much
along the lines of \cite{vp2,vp3,vp4,vp5}, while, the re-summed
perturbation theory used in this paper is closer in spirit
to\cite{vp1}. As mentioned previously, it would be indeed remarkable
if these results can be applied to $D=2+1$ Yang-Mills theories with
or without supersymmetry. We hope to return to this problem in a
future publication.

Finally, it is worth mentioning, that membrane dynamics can also be
used to compute certain glueball masses for gauge theories in spaces
of finite volume\cite{gab}. Since, we can go beyond the computation
of the masses and organize the computation of higher excited states
of  Bosonic membrane theories as quantum spin chain computations,
our results might be applicable to improve upon the glueball
spectroscopy of gauge theories in spaces of  finite volume. \\
{\bf Acknowledgements:} We are deeply indebted to V.P.Nair for many
enlightening discussions, constant encouragement  and for his
collaboration during the initial stages of this work. It is a great
pleasure to thank J.Dai, D.Karabali, A.P.Polychronakos and
S.G.Rajeev for many useful discussions and for their comments on an
earlier version of this manuscript. The author also wishes to thank
T.Klose for discussions about the integrability of the BMN matrix
model and G.Mandal for correspondences about the Myers' model.
\section{Appendix A:} In this appendix we
outline the Bethe equations for an $su(n)$ invariant spin chain with
Hamiltonian given by\beq H = -\sum_l P_{l,l+1}. \eeq This discussion
is meant to provide the background material for various formulae
related to Bethe ansatzae throughout the paper. For a comprehensive
discussion  of $su(n)$ Bethe ansatz techniques we shall refer to the
original papers\cite{genh1,genh2,genh3,genh4}.

The spins or the relevant chain are taken to be in the $(n)$
representation of the algebra, and hence have $n$ components. To
find the spectrum of the Hamiltonian, one starts with the Lax
operator that acts on $\mathcal{H}_l\otimes \mathcal{V}_a$.
$\mathcal{H}_l$ is the 'one particle Hilbert space' associated with
the site $l$ and it is nothing but $\mathcal{C}^n$. $\mathcal{V}_a$
is the auxiliary vector space, and as a vector space, it is also
$\mathcal{C}^n$. The Lax operator is\cite{genh4}\beq L_{l,a} =
a(\mu) I_{l,a}+ b(\mu)P_{l,a}\eeq where\beq a(\mu ) =\frac{\mu +
i/2}{\mu - i/2}, b(\mu) = -\frac{i/2}{\mu - i/2}.\eeq One builds the
transfer matrix from the Lax operator as\beq T_{J,a}(\mu )
=L_{J,a}\cdots L_{1,a}.\eeq This transfer matrix satisfies the
Yang-Baxter algebra \beq \mathcal{R}_{ab}(\mu - \nu)
T_{J,a}(\mu)T_{J,b}(\nu) = T_{J,b}(\nu) T_{J,a}(\mu)
\mathcal{R}_{ab}(\mu - \nu). \eeq The trace of the transfer matrix
over the auxiliary space $t(\mu) = \mbox{Tr}_aT_{J,a}(\mu)$ is the
generator of conserved commuting charges, i.e.  \beq [t(\mu),
t(\nu)] = 0.\eeq The conserved charges can be obtained in the
standard fashion by expanding the transfer matrix around $\mu =
-\frac{i}{2}$. \beq \mathcal{O}^l = i\left(\frac{d}{d\mu}\right)^l
\ln[t(\mu)t(0)^{-1}]|_{\mu =-\frac{i}{2}}.\eeq In particular, \beq H
= \mathcal{O}^1 - J.\eeq The momentum is given, in terms of the
Bethe roots, by\beq e^{-ip} = \frac{\mu - i/2}{\mu + i/2},\eeq while
the dispersion relation \beq \epsilon(p) = -2\cos(p) \eeq
becomes\beq \epsilon(\mu ) = -2\left(\frac{\mu ^2 -1/4}{\mu^2 +
1/4}\right).\eeq

The Bethe equations are\beq \left( \frac{u_{m,i} + i\vec{\alpha }
_{m}. \vec{w}}{u_{m,i} - i\vec{\alpha }_{m}.\vec{w}}\right)^J=
\prod_{j\neq i}^{n_m}\frac{u_{m,i} -u_{m,j}+ i\vec{\alpha } _{m}.
\vec{\alpha}_m}{u_{m,i} -u_{m,j}- i\vec{\alpha } _{m}.
\vec{\alpha}_m}\prod_{m'\neq m}\prod_{j\neq i}^{n_{m'}}\frac{u_{m,i}
-u_{m',j}+ i\vec{\alpha } _{m}. \vec{\alpha}_{m'}}{u_{m,i}
-u_{m',j}- i\vec{\alpha } _{m}. \vec{\alpha}_{m'}}\eeq $\alpha _m$
are the simple roots of the Lie algebra and $w$ is the highest
weight of the representation. The eigenvalues of the Hamiltonian in
terms of the solution of the Bethe equations requires further
details of the Bethe ansatz, and we shall simply quote the result
here while referring to\cite{genh1,genh2,genh3,genh4} for a more
complete derivation. The total energy, is given in terms of the
dispersion relation by\beq E = -\sum_{j=1}^m \epsilon(u_{1,j}).\eeq
It is important to note however that only roots of the first type
enter the dispersion relation, while the other roots only affect the
energies indirectly.

To be able to use the Bethe equations, one also needs information
about the roots and weights for the Lie algebra, which we also list
below for the sake of completeness. For $su(n)$, the simple roots
are given by \beq \alpha _m = \nu_m - \nu_{m+1}, m = 1\cdots
n-1,\eeq while the $n-1$ dimensional weights are given by \beq
[\nu_j]_m = \frac{1}{\sqrt{2m(m+1)}}\left(\sum_{k=1}^m \delta_{j,k}
- m\delta_{j,m+1}\right).\eeq The weights satisfy \beq
\nu_m.\nu_{m'} = -\frac{1}{2n} + \frac{1}{2}\delta_{mm'},\eeq while
\beq \alpha _m.\alpha_{m'} = \delta_{mm'} -
\frac{1}{2}\delta_{m,m'+1}.\eeq The heighest weight in the defining
representation $\vec{w} = \vec{\nu}_1$.

\end{document}